\documentclass[aps,pra,twocolumn,showpacs,superscriptaddress]{revtex4}
\usepackage{graphicx}
\usepackage{amsmath}
\usepackage{amssymb}
\usepackage{lscape}

\input{epsf}

\begin{document}

\title{Spin structure of harmonically trapped one-dimensional atoms with spin-orbit coupling}

\author{Q. Guan}
\affiliation{Department of Physics and Astronomy,
Washington State University,
  Pullman, Washington 99164-2814, USA}
\author{D. Blume}
\affiliation{Department of Physics and Astronomy,
Washington State University,
  Pullman, Washington 99164-2814, USA}

\date{\today}

\begin{abstract}
We introduce a theoretical approach to determine the spin structure of harmonically
trapped atoms with two-body zero-range interactions subject to an
equal mixture of Rashba and Dresselhaus spin-orbit coupling created through
Raman coupling of atomic hyperfine states. The spin structure of bosonic and fermionic two-particle systems 
with finite and infinite two-body interaction strength $g$ is calculated. 
Taking advantage of the fact that the $N$-boson and $N$-fermion systems with infinitely
large coupling strength $g$ are analytically solvable for vanishing spin-orbit coupling strength $k_{so}$
and vanishing Raman coupling strength $\Omega$, we develop an effective spin model that is accurate to second-order 
in $\Omega$ for any $k_{so}$ and infinite $g$. 
The three- and four-particle systems are considered explicitly.
It is shown that the effective spin Hamiltonian, 
which contains a Heisenberg exchange term and an anisotropic Dzyaloshinskii-Moriya exchange term, 
describes the transitions 
that these systems undergo with the change of $k_{so}$ as a competition between 
independent spin dynamics and nearest-neighbor spin interactions. 
\end{abstract}

\maketitle

\section{Introduction}
\label{sec_introduction}
Spin-orbit (more precisely, spin-momentum) coupled systems continue to attract a great deal of attention due to 
the rich physics of the spin-Hall effect, topological insulators, and Majorana fermions~\cite{galitski and spielman,hasan,kato,majorana,rev huizhai}.
While these topics fall traditionally into condensed matter territory, 
recent advances in cold atomic gases have led to a fruitful cross fertilization of atomic and condensed matter physics.
On the experimental side, synthetic gauge fields have been realized in ultracold atom systems~\cite{6,7,8,9,10,11,12,13,14,15}.
On the theoretical side,
systems with different kinds of spin-orbit coupling have been studied at the many- and few-body levels~\cite{mean field1, mean field2, mean field3,6, q guan,x y yin,greene,luu,ho,zhang}.
Understanding the effects of spin-orbit coupling in few-atom systems opens the door to a bottom-up understanding of many-body systems. 
Nowadays, ultracold few-atom systems can be prepared and manipulated in experiments~\cite{jochim,bloch}.
For example, taking advantage of a Fano-Feshbach resonance, 
the interaction between the atoms can be tuned~\cite{chin}.
Moreover, the confinement geometry can be changed from three-dimensional to effectively two-dimensional to effectively one-dimensional~\cite{dalfovo}.
These experimental advances were guided by and stimulated a good number of theoretical few-body studies~\cite{Olshanii,24,25,girardeau,busch,28,29,30,31,32,33,34,35,36}.
Much analytical work has been done by approximating the true alkali atom-alkali atom potential by a 
zero-range contact potential~\cite{busch,28,29,30,33,34,35,36,37,38,39}. 
This approximation captures the low energy regime reliably but fails to reproduce high-energy properties (such as the characteristics of deeply bound states).
Assuming zero-range interactions, 
the energy spectra of two one-dimensional bosons and two one-dimensional fermions with arbitrary two-body coupling constant and spin-orbit coupling strength have 
been calculated and
the interplay between 
the spin-orbit coupling term, Raman coupling term, and the two-body potential has been analyzed~\cite{q guan}.

Extending earlier work~\cite{q guan,ho}, 
this paper investigates the spin structure of harmonically
trapped atoms with two-body zero-range interactions subject to an
equal mixture of Rashba and Dresselhaus spin-orbit coupling created through
Raman coupling of atomic hyperfine states.
The spin structure of two identical one-dimensional bosons and two identical one-dimensional fermions 
with finite and infinite two-body interaction strength $g$ is calculated 
for weak to strong spin-orbit coupling strength.
The bosonic and fermionic systems display, in general, different behaviors as the 
spin-orbit coupling strength $k_{so}$ and the two-body
coupling constant $g$ are changed.
For infinite $g$, however, the bosonic and fermionic systems display the same spin structure.
An interesting transition of the spin structure is found in going from small to large $k_{so}$.
To understand the system behaviors for finite and infinite $g$,
an effective Hamiltonian for any $k_{so}$ that is accurate up to second order 
in the Raman coupling strength $\Omega$ is derived.
Effective Hamiltonian have been utilized in various areas of physics and 
the Hubbard model~\cite{hubbard}, the Born-Oppenheimer approximation~\cite{born}, 
the Ising model~\cite{Ising}, and the Heisenberg spin chain~\cite{heisenbergchain} are 
prominent examples.  
Various approaches to generate effective Hamiltonian and the connection between perturbation theory
and selected effective Hamiltonian have been discussed in Refs.~\cite{eff1,eff2,eff3,eff4,eff5,eff6}.

For infinitely large $g$ and arbitrary $N$, 
we integrate out the spatial degrees of freedom and recast the resulting effective Hamiltonian
in terms of spin operators. Single spin terms are proportional to $\Omega$ 
while spin-spin interactions are proportional to $\Omega^{2}$.
Effective spin Hamiltonian have been derived previously for one-dimensional
systems without spin-orbit and Raman coupling~\cite{H pu,artem,Levinsen,Deuretzbacher}.
In those cases, spin-spin interactions were introduced by allowing for 
small deviations from $|g|=\infty$; in essence,
this introduces a tunneling term.
In our case, the single spin and the spin-spin terms are introduced by the Raman coupling.
The single spin term is proportional to $\Omega$ and has been discussed in Ref.~\cite{ho}.
In essence, the Raman coupling creates an effective local $\vec{B}$-field that the spins follow.
The spin-spin interaction term has, to the best of our knowledge, not been discussed before in this context.
This term contains two contributions. 
The first contribution is of the type of the ``usual'' Heisenberg exchange term~\cite{39},
i.e., it is a $\vec{\sigma}_j\cdot\vec{\sigma}_k$ term, 
where $\vec{\sigma}_j$ is the spin operator of spin $j$.
The second contribution is of the type of the anisotropic Dzyaloshinskii-Moriya exchange term,
i.e., it is a one-dimensional analog of the $\vec{D}\cdot(\vec{\sigma}_j\times\vec{\sigma}_k)$ term~\cite{Dzyaloshinskii, Moriya, Keffer},
where $\vec{D}$ is a constant vector.

The remainder of this paper is organized as follows:
Section~\ref{sec_system hamiltonian} defines the system Hamiltonian, 
discusses its symmetries and introduces an effective low-energy Hamiltonian.
Section~\ref{sec_two_particle} 
determines the spin structure of the two-particle system for different $g$ and $k_{so}$.
Using the effective low-energy Hamiltonian,
Sec.~\ref{sec_spin hamiltonian} derives an effective spin Hamiltonian for infinitely large $g$
and applies this Hamiltonian to determine the spin correlations of the three- and four-particle systems.
A transition from a regime where the dynamics is governed by single spin physics to a regime 
where the dynamics is governed by spin-spin interactions is obtained.
Finally, Sec.~\ref{sec_conclusion} summarizes and concludes. 
The appendices contain a number of technical details, including the evaluation of the matrix elements for infinite $g$,
which are needed to construct both the effective Hamiltonian and the full Hamiltonian.
The analytical techniques and results are expected to be useful for other one-dimensional studies. 

\section{System Hamiltonian and general considerations}
\label{sec_system hamiltonian}
We consider $N$ one-dimensional atoms of mass $m$ in a harmonic trap with 
angular trapping 
frequency $\omega$ and zero-range two-body interactions $g\delta(x_j-x_k)$, where $x_j$ is the position 
of the $j$th particle with respect to 
the
center of the trap.
We assume that each particle feels an 
equal mixture of Rashba and Dresselhaus spin-orbit coupling of strength $k_{so}$ 
and Raman coupling of strength $\Omega$. The system Hamiltonian $\tilde{H}$ reads
\begin{equation}
\tilde{H}=H_{sr}\hat{I}+\sum_{j=1}^{N}\frac{\hbar k_{so}}{m}p_{j}\sigma_{y}^{(j)}+\frac{\Omega}{2}\sigma_{x}^{(j)},
\label{H_tilde}
\end{equation}
where
\begin{equation}
H_{sr}=\sum_{j=1}^{N}-\frac{\hbar^2}{2 m}\frac{\partial^2}{\partial x_{j}^{2}}+\frac{1}{2} m \omega^2 x_{j}^2+\sum_{j<k}g\delta\left(x_{j}-x_{k}\right).
\label{H_sr}
\end{equation}
In Eq.~\eqref{H_tilde}, $\hat{I}$ is the $2\times 2$ identity matrix, 
$\sigma_{x}^{(j)}$ and $\sigma_{y}^{(j)}$ are spin-$1/2$ Pauli matrices of the $j$th particle,
and $p_{j}$ is the momentum operator of the $j$th particle along the $x$-direction.
Our goal in the following is to determine the eigenstates $\tilde{\Psi}$ and eigenenergies $\tilde{E}$ of $\tilde{H}$.

To this end, we perform a unitary transformation and define $H=U^{\dagger}\tilde{H}U$, where $U=\Pi_{j=1}^{N}\exp(-i k_{so} x_{j}\sigma_{y}^{(j)})$. 
The transformed Hamiltonian $H$ reads 
\begin{equation}
H=H_{0}\hat{I}+\frac{\Omega}{2}V_{R}
\label{H}
\end{equation}
with
\begin{equation} 
H_{0}=H_{sr}-\frac{N\hbar^2 k_{so}^2}{2 m}
\label{H_0}
\end{equation}
and
\begin{equation}
V_{R}=\sum_{j=1}^{N}\exp(i k_{so} x_{j}\sigma_{y}^{(j)})\sigma_{x}^{(j)}\exp(-i k_{so} x_{j}\sigma_{y}^{(j)}).
\label{V_R}
\end{equation}
The eigenvalues of $H$ are the same as those of $\tilde{H}$ while 
the eigenstates $\Psi$ of $H$ are related to the eigenstates $\tilde{\Psi}$ of $\tilde{H}$ by $\Psi=U\tilde{\Psi}$.
Our strategy for solving the Schr\"odinger equation $H\Psi=E\Psi$ is as follows:
We first find the eigenstates of $H_{0}$ and $H_{0}\hat{I}$ and then account for the Raman coupling term $V_{R}$
through either a matrix diagonalization or an effective low-energy Hamiltonian approach that is accurate to second order in $\Omega$.  

We denote the eigenstates and eigenenergies of $H_{sr}$ by $\phi_n(\vec{x})$ and $E_n$,
where $n$ collectively denotes the quantum numbers needed to label the states and 
$\vec{x}$ collectively denotes the $N$ spatial degrees of freedom, $\vec{x}=(x_1,x_2,\cdot\cdot\cdot,x_N)$.
In general, the $\phi_{n}(\vec{x})$ are not known. 
Since $H_{0}$ is independent of spin, 
the eigenstates of $H_{0}\hat{I}$ can be written as
\begin{equation}
\psi_{n,s_{1},s_{2},\cdot\cdot\cdot,s_{N}}=\phi_{n}(\vec{x})|s_{1},s_{2},\cdot\cdot\cdot,s_{N}\rangle_{y},
\label{eigenstate_H_0}
\end{equation}
where the $|s_{1}, s_{2},\cdot\cdot\cdot,s_{N}\rangle_{y}$ are eigenstates of the operator 
$\sigma_y=\sum_{j=1}^{N}\sigma_{y}^{(j)}$, i.e., $\sigma_y^{(j)}|s_j\rangle_y=s_j|s_j\rangle_y$ with $s_j=\pm1$. 
The eigenenergies of these eigenstates are $E_{n}-N\hbar^2k_{so}^2/(2m)$. 
The eigenstates with the same $\phi_n(\vec{x})$ but different spin configurations are degenerate.
Thus one can form linear combinations of the $\psi_{n,s_{1},s_{2},\cdot\cdot\cdot,s_N}$ such that the resulting eigenstates 
are eigenstates of the $\vec{\sigma}^2$ operator, where $\vec{\sigma}^2=(\sum_{j=1}^N\vec{\sigma}^{(j)})^2$.
For non-vanishing $\Omega$, 
$H$ no longer commutes with $\sigma_y$ or $\vec{\sigma}^2$, 
implying that the total spin and the total projection quantum numbers are no longer conserved.  
However, $H$ still contains symmetries. 
Specifically, 
$H$ commutes with all $P_{jk}\ (j,k=1,...,N)$ operators and the $Y$ operator. 
$P_{jk}$ exchanges both the spatial and spin coordinates of particles $j$ and $k$. 
It follows that the eigenstates of $H$ 
can be classified according to whether or not they change 
sign under the $P_{jk}$ operator. Throughout this paper, we restrict ourselves to the $N$ identical 
boson and $N$ identical fermion sectors.  
The operator
$Y$ can be written as $\Pi_{j=1}^{N}Q_{j}\bigotimes\sigma_{x}^{(j)}$, 
where $Q_{j}$ changes the sign of the spatial coordinate of particle $j$. 
It follows that the eigenstates of $H$, which are also eigenstates of $Y$,
always include an equal mixture of $|s_{1},s_{2},\cdot\cdot\cdot,s_{N}\rangle_{y}$ and $|\bar{s}_1,\bar{s}_2,\cdot\cdot\cdot,\bar{s}_N\rangle_{y}$, 
where $|\bar{s}_j\rangle_{y}$ denotes the state obtained by operating with $\sigma_{x}^{(j)}$ onto $|s_{j}\rangle_{y}$.
Correspondingly, the 
expectation value of $\sigma_y$, and thus $S_y$, 
is always zero.
We label the symmetries of the eigenstates by $(a,b)$,
where $(a,b)=(1,1), (1,-1), (-1,1),$ and $(-1,-1)$ indicate bosonic $(a=1)$ 
or fermionic symmetry $(a=-1)$ with positive $(b=1)$ or negative $(b=-1)$ eigenvalue 
of $Y$.

In what follows, we are interested in the 
expectation values of the
local spins
$S_x(x)$ and $S_z(x)$
in the $x$- and $z$-directions,
\begin{equation}
S_{x}(x)=\sum_{j=1}^{N}\frac{\hbar}{2}\sigma_{x}^{(j)}\delta(x-x_{j})
\label{S_x}
\end{equation}
and
\begin{equation}
S_{z}(x)=\sum_{j=1}^{N}\frac{\hbar}{2}\sigma_{z}^{(j)}\delta(x-x_{j}).
\label{S_z}
\end{equation}
As already mentioned, we have $\langle S_y(x) \rangle=0$.
To quantify the correlations of the system, we define the projector $P_{|M_s|=m}$,
\begin{equation}
P_{|M_{s}|=m}=\sum_{|M_s|=m}|s_1,s_2,\cdot\cdot\cdot,s_N\rangle_y{}_y\langle s_1,s_2,\cdot\cdot\cdot,s_N|,
\label{P_Ms}
\end{equation}
where $M_s$ is the spin projection quantum number in the $y$-direction,
$M_s=s_1+s_2+\cdot\cdot\cdot+s_N$.
The expectation value of $P_{|M_s|=m}$ yields the probability to find the system in the space spanned by all the states with the same absolute value of $M_s$.
For example, for the three particle system, $M_s=\pm 1$ and $\pm 3$.
In this case, $P_{|M_s|=1}$ and $P_{|M_s|=3}$ measure the configurations with $M_s=\pm 1$ and $M_s=\pm 3$, respectively.
 
To diagonalize $H$ directly, we start with the eigenstates of $H_0\hat{I}$ [see Eq.~\eqref{eigenstate_H_0}].
By taking appropriate linear combinations of $\psi_{n,s_1,s_2,\cdot\cdot\cdot,s_N}$, 
the Hamiltonian matrix for the different $(a,b)$ symmetry channels can be set up and diagonalized separately.
The off-diagonal matrix elements of $H$ contain spatial integrals of the form 
$\left(\Omega/2\right)\int_{-\infty}^{\infty}\phi_n^*(\vec{x})\phi_m(\vec{x})e^{\pm 2i k_{so}x_{j}}d\vec{x}$, 
which is the Fourier transform with respect to the coordinate $x_j$ of the product of two eigenstates of $H_0$.
As discussed in Appendix~\ref{appendix_A}, these integrals can be evaluated up to any precision numerically and in some cases can be calculated fully analytically.

Since we are primarily interested in the lowest eigenenergy and eigenstate of $H$ for non-zero $\Omega$,
we derive an effective low-energy Hamiltonian, which describes much of the system dynamics 
accurately and is numerically more tractable.
The idea is to divide the Hilbert space spanned by $H_0\hat{I}$ into two pieces called
$H_L$ and $H_H$~\cite{eff1,eff2}.
The low-energy space $H_L$ contains the eigenstates with energy less than a preset value and the high-energy
space $H_H$ contains the eigenstates with energy larger than this preset value.
The eigenstates of the effective Hamiltonian $H_{eff}$, which accounts for the Raman coupling term
$(\Omega/2)V_R$ perturbatively, are linear combinations of the unperturbed states of $H_0\hat{I}$ that lie in $H_L$. 
We write 
\begin{equation}
H_{eff}=H_{eff}^{(0)}+H_{eff}^{(1)}+H_{eff}^{(2)},
\label{H_eff}
\end{equation}
where 
\begin{eqnarray}
H_{eff}^{(m)}=\sum_{\substack{s_{1},s_{2},\cdot\cdot\cdot,s_{N}\\s_{1}^{'},s_{2}^{'},\cdot\cdot\cdot,s_{N}^{'}\\l_{j},l_{k}\in H_{L}}}|\psi_{l_{j},s_{1},s_{2},\cdot\cdot\cdot,s_{N}}\rangle\nonumber\times\\
\langle\psi_{l_{j},s_{1},s_{2},\cdot\cdot\cdot,s_{N}}|
A_{l_jl_k}^{(m)}|\psi_{l_{k},s_{1}^{'},s_{2}^{'},\cdot\cdot\cdot,s_{N}^{'}}\rangle\langle\psi_{l_{k},s_{1}^{'},s_{2}^{'},\cdot\cdot\cdot,s_{N}^{'}}|
\label{H_eff_m}
\end{eqnarray}
with
\begin{equation}
\label{A_0}
A_{l_jl_k}^{(0)}=H_0,
\end{equation}
\begin{equation}
\label{A_1}
A_{l_jl_k}^{(1)}=\frac{\Omega}{2}V_R,
\end{equation}
and
\begin{widetext}
\begin{eqnarray}
\label{A_2}
A_{l_jl_k}^{(2)}=\frac{\Omega^2}{8}
\sum_{\substack{s_{1}^{''},s_{2}^{''},\cdot\cdot\cdot,s_{N}^{''}\\h_{n} \in H_{H}}}\Bigg(\frac{V_{R}|\psi_{h_{n},s_{1}^{''},s_{2}^{''},\cdot\cdot\cdot,s_{N}^{''}}\rangle
\langle\psi_{h_{n},s_{1}^{''},s_{2}^{''},\cdot\cdot\cdot,s_{N}^{''}}|V_{R}}{E_{l_{j}}-E_{h_{n}}}
+\frac{V_{R}|\psi_{h_{n},s_{1}^{''},s_{2}^{''},\cdot\cdot\cdot,s_{N}^{''}}\rangle
\langle\psi_{h_{n},s_{1}^{''},s_{2}^{''},\cdot\cdot\cdot,s_{N}^{''}}|V_{R}}{E_{l_{k}}-E_{h_{n}}}\Bigg).
\end{eqnarray}
\end{widetext}
\begin{figure}
\vspace*{+0cm}
\hspace*{+0cm}
\includegraphics[width=0.3\textwidth]{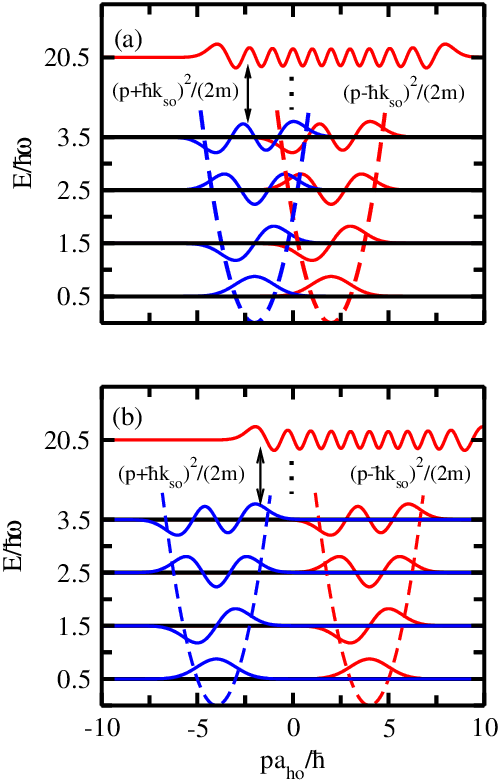}
\vspace*{0cm}
\caption{Single-particle momentum space potentials $(p\pm\hbar k_{so})^2/(2m)$ (dashed lines)
and non-interacting single-particle momentum space wave functions (solid lines)
for (a) $k_{so}a_{ho}=2$ and (b) $k_{so}a_{ho}=4$, respectively.
The potential and wavefunctions centered at $p=-\hbar k_{so}$ are for the spin-up component and those 
centered at $p=\hbar k_{so}$ are for the spin-down component.
For small $k_{so}$, the spin-up and spin-down single-particle wavefunctions in the Hilbert space $H_L$ (here, states with $E\leq 3.5\hbar\omega$)
have overlap, indicating 
that the first-order effective Hamiltonian dominates.
For large $k_{so}$, the spin-up and spin-down single particle wavefunctions in the Hilbert space $H_L$ 
have essentially zero overlap, indicating that 
coupling to highly excited states is dominant, leading to a dominant second-order effective Hamiltonian.   
}
\label{intuitive}
\end{figure}
An important point is that the second-order term $A_{l_jl_k}^{(2)}$ runs over all states from 
the $H_H$ 
Hilbert space. In essence, this leads to a renormalization of the matrix elements in the 
low-energy space $H_L$. 

To obtain an intuitive understanding of the effective Hamiltonian
approach, Figs.~\ref{intuitive}(a) and~\ref{intuitive}(b) show the single-particle momentum space potentials
$(p\pm\hbar k_{so})^2/(2m)$ for $k_{so}a_{ho}=2$ and $k_{so}a_{ho}=4$, respectively. 
In this picture, the potentials centered at $p=-\hbar k_{so}$ and $p=\hbar k_{so}$
are occupied by the spin-up and spin-down components of the wave function. The Raman coupling term 
introduces a mixing of the spin-up and spin-down potentials. 
Within the effective Hamiltonian approach, this is accounted for by $H_{eff}^{(1)}$ and $H_{eff}^{(2)}$.
To estimate the relative importance of $H_{eff}^{(1)}$ and $H_{eff}^{(2)}$, 
we consider the $g=\infty$ case (this case is discussed in detail in Sec.~\ref{sec_spin hamiltonian}).
For $g=\infty$, the $N$-particle wave functions are constructed from the non-interacting single-particle
harmonic oscillator states. 
For $N=4$, e.g., the single-particle states shown in Figs.~\ref{intuitive}(a) and~\ref{intuitive}(b) contribute. 
The Hamiltonian $H_{eff}^{(1)}$ couples states centered at $p=-\hbar k_{so}$ and $\hbar k_{so}$. 
Visually, it is clear that the coupling (i.e., the overlap between the different single-particle states centered at $p=-\hbar k_{so}$ and $\hbar k_{so}$)
decreases with increasing $k_{so}$. 
As a consequence, the Hamiltonian term $H_{eff}^{(2)}$ may carry a higher ``weight'' than $H_{eff}^{(1)}$ 
for sufficiently large $k_{so}$ despite the fact that $H_{eff}^{(2)}$ is suppressed by the factor $\Omega/(\hbar\omega)$ for $\Omega<\hbar\omega$.
More quantitatively, the $N$th single-particle state is distributed in the momentum interval $(\pm\hbar k_{so}-\hbar k_{F}, \pm\hbar k_{so}+\hbar k_{F})$,
where $k_{F}a_{ho}\sim\sqrt{2N}$. Thus for $k_{so}\ge\sqrt{2N}/a_{ho}$, the coupling matrix elements entering into $H_{eff}^{(1)}$ are expected to be small
while the coupling matrix elements entering into $H_{eff}^{(2)}$ are expected to have a comparatively large amplitude 
[the vertical arrow in Figs.~\ref{intuitive}(a) and~\ref{intuitive}(b) indicates
the momentum region where the $N$th single-particle state overlaps with a highly excited single-particle state].
Our pictorial analysis is confirmed by our quantitative calculations. 
Specifically, as detailed in Sec.~\ref{subsection_N2-4}, the spin structure changes dramatically at a critical $k_{so}$ value,
beyond which the term $H_{eff}^{(2)}$ dominates over the term $H_{eff}^{(1)}$. Analogous arguments can be made for finite $g$.
 
Based on the effective Hamiltonian given in Eq.~\eqref{H_eff}, the next section calculates the spin structure for two-particle systems 
with arbitrary
coupling constant $g$ and compares the results with those obtained from the full (``brute force'') diagonalization.

\section{Two-particle system with arbitrary $g$}
\label{sec_two_particle}
 For the two-particle system, 
the Schr\"odinger equation for $H_{sr}$ has been solved analytically for 
arbitrary two-body interaction strength in the literature~\cite{busch}. 
The eigenstates are $\phi_{n}(\vec{x})=\Phi_{p}(R)\varphi_{q}(r)$ with eigenenergy $E_{n}=(p+2q+1)\hbar\omega$, 
where the center of mass coordinate $R$ and the
relative coordinate $r$ are defined through $R=(x_1+x_2)/\sqrt{2}$ 
and $r=(x_1-x_2)/\sqrt{2}$.
The center of mass eigenstates $\Phi_{p}(R)$ are the harmonic oscillator eigenstates with energy $(p+1/2)\hbar\omega$,
where $p=0,1,\cdot\cdot\cdot$. 
For states with even parity in the relative coordinate, 
the relative eigenstates
$\varphi_{q}(r)$ are given by $N_{q}U(-q,1/2,(r/a_{ho})^2)\exp[-r^2/(2a_{ho}^2)]$,
where $a_{ho}$ denotes the harmonic oscillator length, 
\begin{equation}
a_{ho}=\sqrt{\frac{\hbar}{m\omega}},
\end{equation} 
$N_{q}$ is a normalization constant and $U$ is the confluent hypergeometric function. 
In this even parity case, $q$ denotes a non-integer quantum number, 
which is determined by the transcendental equation~\cite{busch}
\begin{equation}
\frac{2\Gamma(-q+1/2)}{\Gamma(-q)}=-\frac{g}{\sqrt{2}\hbar\omega a_{ho}}.
\label{trans}
\end{equation}
For states with odd parity in the relative coordinate, 
the relative eigenstates $\varphi_{q}(r)$ are again harmonic oscillator eigenstates with eigenenergy $(2q+1/2)\hbar\omega$,
where $q$ takes half-integer values, i.e., $q=1/2,3/2,\cdot\cdot\cdot$.
Since the two-body $\delta$-function only acts at $r=0$,
the odd-parity eigenstates and eigenenergies are independent of $g$. 
The eigenenergies of the even-parity relative eigenstates, in contrast, 
change with $g$. 
Specifically, the energy $(2q+1/2)\hbar\omega$ increases with increasing $g$.
For infinite $g$,
the ground state of the two-particle system is doubly degenerate,
i.e., the relative energies 
of the even-parity state and the odd-parity state coincide, 
yielding a total energy of $2\hbar\omega$.
This degeneracy plays, as we show below, an important role when the spin-orbit and Raman coupling 
strengths are non-zero.
 
When $k_{so}$ and $\Omega$ are non-zero, the center of mass and relative degrees of 
freedom are coupled. In order to describe the interplay between the two-body interaction and the Raman and spin-orbit couplings, 
the states with spatial parts $\Phi_0(R)\varphi_{q_{0}}(r)$ and $\Phi_0(R)\varphi_{q_{1}}(r)$ span the $H_{L}$ space.
Here $\varphi_{q_{0}}$ and $\varphi_{q_{1}}$ denote the energetically lowest-lying relative even-parity and odd-parity states, respectively. 
We choose the spatial basis functions to be real. Our motivation for choosing this $H_L$ space
is as follows. 
We want the resulting effective low-energy Hamiltonian to describe the ground state of the two-particle system
with good accuracy for all $g$. 
For small $g$, e.g., the low-energy Hamiltonian constructed using $\Phi_{0}(R)\varphi_{q_0}(r)$ would suffice 
since the energy of the state $\Phi_{0}(R)\varphi_{q_1}(r)$ is close to $\hbar\omega$ higher in energy.
Including this state in $H_L$ changes the effective low-energy Hamiltonian and its resulting eigenenergies negligibly.
For large $g$, in contrast, the state $\Phi_{0}(R)\varphi_{q_1}(r)$ needs to be included in $H_L$ 
since its energy is, as discussed above, nearly degenerate with the state $\Phi_{0}(R)\varphi_{q_0}(r)$.
Thus, the space $H_L$ identified above is the minimal space needed if an effective description of 
the ground state for all $g$ (small and large) is sought.

For two identical bosons, the ground state $\psi_{gr}^b$ of the effective Hamiltonian $H_{eff}$ is 
\begin{eqnarray}
\psi_{gr}^{b}=\frac{C_{1}^{b}}{\sqrt{2}}\Phi_{0}(R)\varphi_{q_{0}}(r)(|\uparrow\uparrow\rangle_{y}+|\downarrow\downarrow\rangle_{y})\nonumber\\
+\frac{C_{2}^{b}}{\sqrt{2}}\Phi_{0}(R)\varphi_{q_{0}}(r)(|\uparrow\downarrow\rangle_{y}+|\downarrow\uparrow\rangle_{y})\nonumber\\
+\frac{C_{3}^{b}}{\sqrt{2}}\Phi_{0}(R)\varphi_{q_{1}}(r)(|\uparrow\downarrow\rangle_{y}-|\downarrow\uparrow\rangle_{y}).
\label{grN2}
\end{eqnarray}
The coefficients $C_1^{b}$-$C_3^{b}$ are obtained by diagonalizing the effective low-energy Hamiltonian $H_{eff}$.
Using $\psi_{gr}^b$ to calculate the expectation values of $S_x(x)$ and $S_z(x)$, we find
\begin{equation}
\label{sxN2}
\langle S_{x}^{b}(x)\rangle=\frac{\hbar}{2}C_{x}^{b}n_{x}^{b}(x)
\end{equation}
and
\begin{equation}
\label{szN2}
\langle S_{z}^{b}(x)\rangle=\frac{\hbar}{2}C_{z}^{b}n_{z}^{b}(x),
\end{equation}
where
\begin{equation}
\label{sxcoefficientN2}
C_{x}^{b}=C_{1}^{b}(C_{2}^{b})^{*}+(C_{1}^{b})^{*}C_{2}^{b},
\end{equation}
\begin{equation}
\label{sxdensityN2}
n_{x}^{b}(x)=\int_{-\infty}^{\infty}\Phi_{0}^{2}(R)\varphi_{q_{0}}^{2}(r)[\delta(x-x_{1})+\delta(x-x_{2})]d\vec{x},
\end{equation}
\begin{equation}
\label{szcoefficientN2}
C_{z}^{b}=i[C_{1}^{b}(C_{3}^{b})^{*}-(C_{1}^{b})^{*}C_{3}^{b}],
\end{equation}
and
\begin{equation}
\label{szdensityN2}
n_{z}^{b}(x)=\int_{-\infty}^{\infty}\Phi_{0}^{2}(R)\varphi_{q_{0}}(r)\varphi_{q_{1}}(r)[\delta(x-x_{1})-\delta(x-x_{2})]d\vec{x}.
\end{equation}

For two identical fermions, the ground state of the effective Hamiltonian is 
\begin{eqnarray}
\label{grN2_f}
\psi_{gr}^{f}=\frac{C_{1}^{f}}{\sqrt{2}}\Phi_{0}(R)\varphi_{q_{1}}(r)(|\uparrow\uparrow\rangle_{y}+|\downarrow\downarrow\rangle_{y})\nonumber\\
+\frac{C_{2}^{f}}{\sqrt{2}}\Phi_{0}(R)\varphi_{q_{1}}(r)(|\uparrow\downarrow\rangle_{y}+|\downarrow\uparrow\rangle_{y})\nonumber\\
+\frac{C_{3}^{f}}{\sqrt{2}}\Phi_{0}(R)\varphi_{q_{0}}(r)(|\uparrow\downarrow\rangle_{y}-|\downarrow\uparrow\rangle_{y}).
\end{eqnarray}
It can be seen that the structure of $\psi_{gr}^f$ is very similar to that of $\psi_{gr}^b$, Eq.~\eqref{grN2}.
The only differences are that the superscript $b$ is replaced by $f$, that $\varphi_{q_0}$ is replaced by $\varphi_{q_1}$ (twice) 
and that $\varphi_{q_1}$ is replaced by $\varphi_{q_0}$ (once).
It follows that the expressions that describe the spin structure of the fermionic system are nearly identical to those that 
describe the spin structure of the bosonic system. 
Specifically, Eqs.~\eqref{sxN2}-\eqref{szdensityN2} apply to the fermionic system if the superscript $b$
is replaced by $f$, $\varphi_{q_0}$ is replaced by $\varphi_{q_1}$, and $\varphi_{q_1}$ is replaced by $\varphi_{q_0}$.

\begin{figure}
\vspace*{0cm}
\hspace*{-0.5cm}
\includegraphics[width=0.43\textwidth]{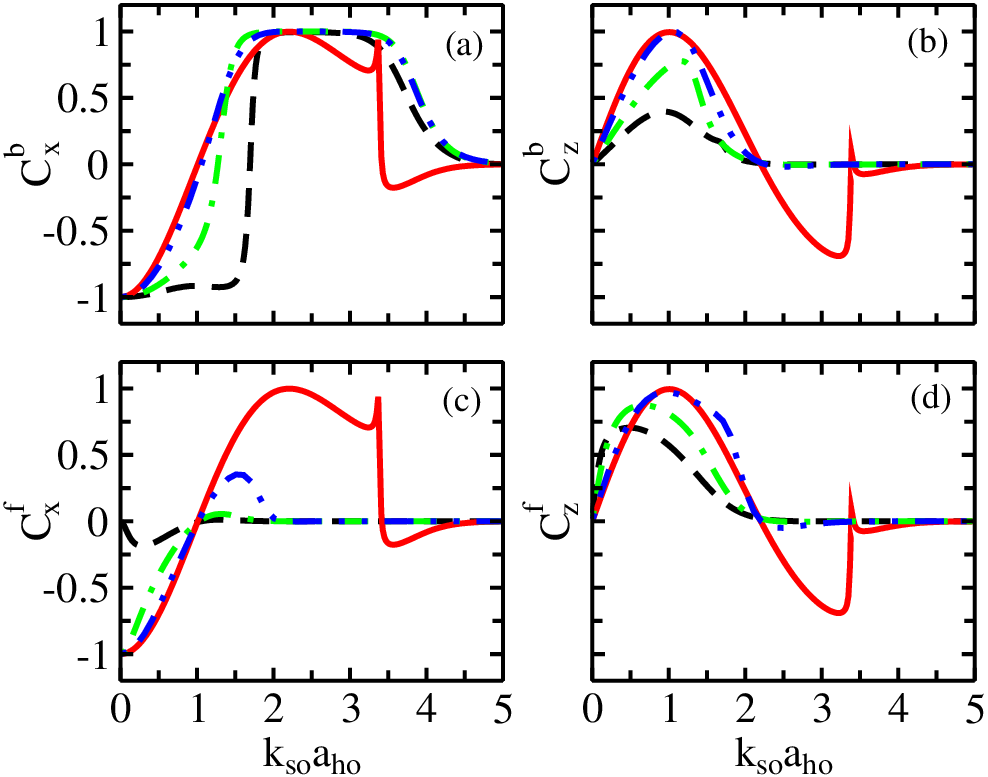}
\vspace*{0cm}
\caption{Spin structure coefficients (a) $C_{x}^b$, (b) $C_{z}^b$, (c) $C_{x}^f$, and (d) $C_{z}^f$
for the $N=2$ ground state obtained within the effective Hamiltonian approach for 
$\Omega=\hbar\omega/2$. 
The spin structure coefficients in (a) and (b) are for the system consisting of two identical bosons.
The spin structure coefficients in (c) and (d) are for the system consisting of two identical fermions.
Dashed, dot-dashed, dot-dot-dashed, and solid lines show the spin structure coefficients for 
$g=\sqrt{2}\hbar\omega a_{ho}, 3\sqrt{2}\hbar\omega a_{ho}, 15\sqrt{2}\hbar\omega a_{ho}$
and $\infty$, respectively.
For infinitely large $g$ (solid lines), the spin structure coefficients display a ``sharp spike'' 
at $k_{so}a_{ho}\approx 3.4$; this is 
the result of an avoided crossing between the ground and first excited 
states.}
\label{2particle_C}
\end{figure}

\begin{figure}
\vspace*{0cm}
\hspace*{-0.5cm}
\includegraphics[width=0.43\textwidth]{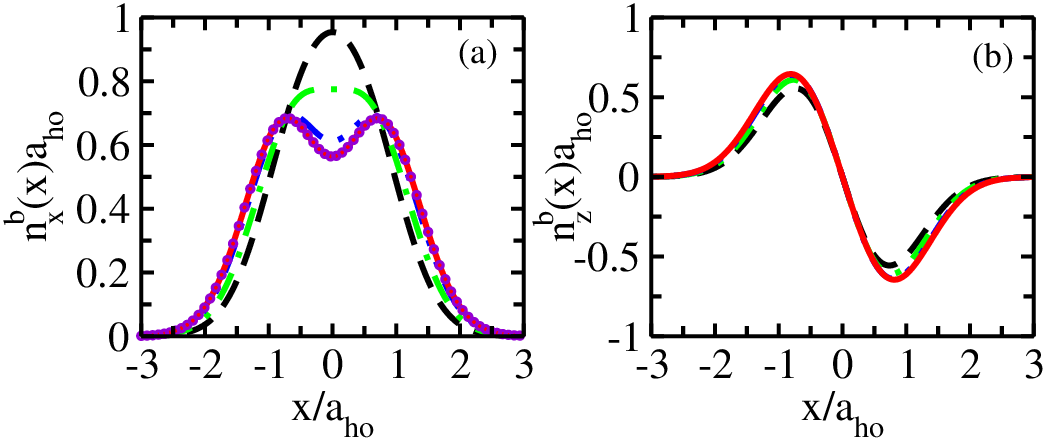}
\vspace*{0cm}
\caption{Spin structure densities for the $N=2$ ground state for 
$\Omega=\hbar\omega/2$ 
and varying $g$. 
Panels (a) and (b) show the spin structure densities $n_x^b(x)$ and $n_z^b(x)$, respectively. 
Dashed, dot-dashed, dot-dot-dashed, and solid lines show the spin densities for 
$g=\sqrt{2}\hbar\omega a_{ho}, 3\sqrt{2}\hbar\omega a_{ho}, 15\sqrt{2}\hbar\omega a_{ho},$
and $\infty$, respectively. 
The spin structure density is independent of $k_{so}$.
The spin structure density $n_x^f(x)$ is independent of $g$ and shown by the circles in (a).
The spin structure density $n_z^f(x)$ coincides with $n_z^b(x)$ for all $g$.}
\label{2particle_n}
\end{figure}

We refer to $C_x^{b/f}$ and $C_z^{b/f}$ as the spin structure coefficients 
and to 
$n_x^{b/f}(x)$ and $n_z^{b/f}(x)$ 
as the spin structure densities.
Note that these densities can be negative; a negative density indicates that the spin points in the negative direction. 
Importantly, the spin structure densities depend on $g$ but are independent of $k_{so}$ and $\Omega$.
The spin structure coefficients, in contrast, depend on $k_{so}, \Omega$ and $g$.
To obtain the spin structure within the effective low-energy Hamiltonian approach,
the spin structure density (see Fig.~\ref{2particle_n}) gets multiplied by the spin structure coefficients,
shown in Fig.~\ref{2particle_C} as a function of $k_{so}$ for various $g$ and fixed $\Omega$ 
($\Omega=\hbar\omega/2$).

In general, the spin structures of the bosonic and fermionic systems differ.
For a weak two-body interaction (small positive $g$) and Raman coupling 
strength $\Omega=\hbar\omega/2$ with vanishing spin-orbit coupling strength,
the coupling between the singlet and the triplet states vanishes.
Since the relative even parity state has a lower energy than the lowest relative odd-parity state for $g<\infty$,
the ground states for two identical bosons and fermions are triplet and singlet states, respectively.
Correspondingly, the coefficient $C_3^b$ in Eq.~\eqref{grN2} and the coefficients $C_1^f$ and $C_2^f$ in Eq.~\eqref{grN2_f} vanish,
yielding 
$\langle S_z^b(x)\rangle =\langle S_x^f(x)\rangle =\langle S_z^f(x)\rangle =0$
and
$\langle S_x^{b}(x)\rangle\neq 0$. 
For a non-vanishing spin-orbit coupling strength, the singlet and triplet states are coupled,
leading to non-zero $\langle S_z^b(x)\rangle, \langle S_x^f(x)\rangle,$ and $\langle S_z^f(x)\rangle$.
With increasing $g$, the energy difference between the singlet and triplet states for $k_{so}=0$ decreases.
As a consequence, for a fixed and relatively small $k_{so}$, the spin structures for two identical bosons and two identical fermions are 
more similar for larger $g$ than for smaller $g$ 
[see Figs.~\ref{2particle_C}(a) and ~\ref{2particle_C}(c) for $k_{so}a_{ho}\lesssim 1.5$].
However, the coupling between the singlet and triplet states is weakened with increasing $k_{so}$.
Specifically, for fixed $g$, the spin structures for bosons and fermions differ more as $k_{so}$ increases 
[see Figs.~\ref{2particle_C}(a) and ~\ref{2particle_C}(c) for $k_{so}a_{ho}>2$].
For a larger $k_{so}$, a larger $g$ is needed to get the triplet (singlet) state mixed significantly into the ground state
for two identical fermions (bosons), resulting in similar spin structures for bosons and fermions.
For an infinitely large $g$, the relative even and odd parity states have the same energy and 
the spin structures for bosonic and fermionic systems are identical for all $k_{so}$ and $\Omega$ (see the solid lines in Fig.~\ref{2particle_C}).
This effect can be attributed to the Bose-Fermi duality (see next section for more details).

\begin{figure}
\vspace*{0cm}
\hspace*{-0.5cm}
\includegraphics[width=0.43\textwidth]{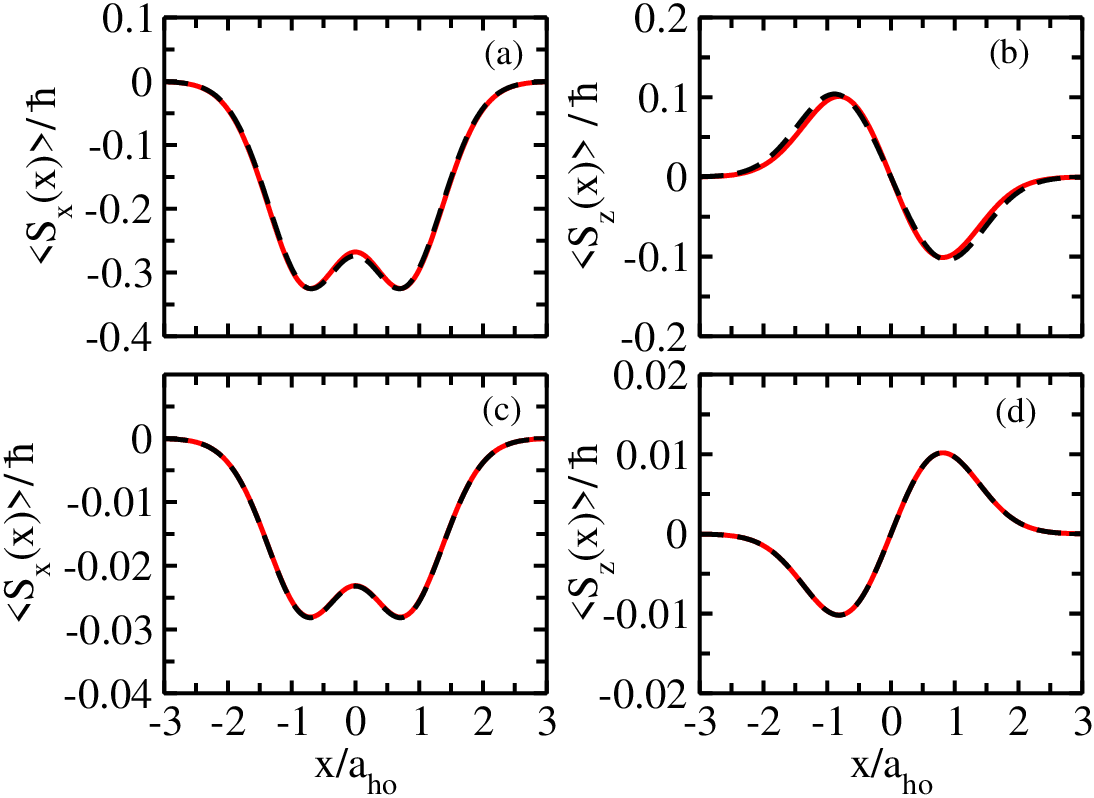}
\vspace*{0cm}
\caption{Benchmarking the effective Hamiltonian approach for the $N=2$ ground state
for 
$\Omega=\hbar\omega/2$ and $g=\infty$. 
Panels (a) and (b) show $\langle S_x(x)\rangle$ and $\langle S_z(x)\rangle$ for 
$k_{so}a_{ho}=1/5$; 
panels (c) and (d) show $\langle S_x(x)\rangle$ and $\langle S_z(x)\rangle$
for $k_{so}a_{ho}=4$. The solid lines show the spin structure obtained from the exact diagonalization
while the dashed lines show the spin structure obtained within the effective Hamiltonian approach.
On the scale shown, the solid and dashed lines nearly coincide.
}
\label{2particle_compare}
\end{figure}

Our discussion so far has been based on the effective Hamiltonian approach. 
To benchmark this approach, we compare the effective Hamiltonian results with 
those obtained from the exact diagonalization for various $g$, $k_{so},$ and $\Omega$
combinations. The agreement is good for all cases considered. 
As an example, Fig.~\ref{2particle_compare} compares the two-particle spin structure for infinite $g$ 
and 
$\Omega=\hbar\omega/2$. 
Figures~\ref{2particle_compare}(a) and~\ref{2particle_compare}(b) 
are for small $k_{so}$ 
($k_{so} a_{ho} \ll 1$)
and Figs.~\ref{2particle_compare}(c) and~\ref{2particle_compare}(d) 
are for large $k_{so}$
($k_{so} a_{ho} \gg 1$).
The agreement between the effective Hamiltonian approach results (solid lines) and
the exact diagonalization approach results (dashed lines) is very convincing.

\section{N-particle system with $g=\infty$}
\label{sec_spin hamiltonian}
\subsection{Formulation}
When the two-body coupling constant $g$ is infinitely large, 
the particles cannot pass through each other. 
In the absence of the spin-orbit and Raman coupling terms,
the atomic gas behaves
like a Tonks-Girardeau gas~\cite{girardeau0,Olshanii}.
The Tonks-Girardeau gas has a large degeneracy and bosons fermionize~\cite{girardeau0,girardeau,girardeau1}.
The fact that the particles cannot pass through each other implies that 
the particles can be ordered.
Since there are $N!$ ways  
to order the particles, 
the degeneracy of each eigenstate of $H_{sr}$ with $g=\infty$ is $N!$
if the particle exchange symmetry is not being enforced.
We thus pursue an approach where we first determine the eigenstates of 
$H_{sr}\hat{I}$ for a fixed particle ordering. The resulting eigenstates are 
then used either to calculate the eigenstates and eigenenergies of $H$ 
through exact diagonalization or to calculate the eigenstates and eigenenergies 
of $H_{eff}$ by identifying an appropriate subspace $H_L$.
The fact that the particles can be ordered also allows us to derive an 
effective spin Hamiltonian $H_{eff}^{spin}$ that is independent of the spatial coordinates.

For infinite $g$, the eigenstates of $H_{sr}$ can be written as~\cite{girardeau}
\begin{equation} 
\phi_{n_{1},n_{2},...,n_{N}}(\vec{x})=D(n_{1},n_{2},\cdot\cdot\cdot,n_{N})\Theta_{x_{j_{1}}<x_{j_{2}}<\cdot\cdot\cdot<x_{j_{N}}},
\label{wavefunc_infg}
\end{equation}
where $D(n_{1},n_{2},\cdot\cdot\cdot,n_{N})$ denotes the Slater determinant constructed from $N$ one-dimensional harmonic oscillator 
eigenfunctions with quantum numbers $n_{1},n_{2},\cdot\cdot\cdot,n_{N}$ ($n_1\neq n_2\neq\cdot\cdot\cdot\neq n_N$). 
The sector
function $\Theta_{x_{j_{1}}<x_{j_{2}}<\cdot\cdot\cdot<x_{j_{N}}}$ is $1$ for $x_{j_{1}}<x_{j_{2}}<\cdot\cdot\cdot<x_{j_{N}}$ and zero otherwise. 
For the ground state, e.g., we have $n_{1}=0,n_{2}=1,\cdot\cdot\cdot,n_{N}=N-1$. 
To construct eigenstates of $H_{sr}\hat{I}$, 
we include---as before---the spin part $|s_{1},s_{2},\cdot\cdot\cdot,s_{N}\rangle_{y}$.
It is important to realize that there is no coupling 
between the eigenstates for different particle orderings, 
i.e., the Raman coupling term $(\Omega/2)V_R$ does not couple states with 
different particle orderings. This implies that the Hilbert space can
be divided into $N!$ independent subspaces.
A state in a given subspace can be mapped onto
a state in a different subspace with the same eigenenergy
through the application of one or more permutation operators.
For example, for the $N=3$ system,
the particle ordering $x_1<x_2<x_3$ can be changed to the
ordering $x_2<x_1<x_3$ through the application
of the $P_{12}$ operator. 
This property is used, after solving the problem in one of the $N!$ subspaces, to construct fully
symmetric bosonic or fully anti-symmetric fermionic eigenstates from the non-symmetrized eigenstates
that span one of the $N!$ distinct Hilbert spaces.

Without loss of generality,
we discuss the ordering $x_1<x_2<\cdot\cdot\cdot<x_N$.
Within this subspace, the evaluation of the Hamiltonian
matrix elements involve integrals of the form 
\begin{eqnarray}
\mathcal{D}_{\substack{n_{1},n_{2}...n_{N}\\n_{1}^{'},n_{2}^{'},\cdot\cdot\cdot,n_{N}^{'}}}^{j}=\int_{-\infty}^{\infty}D(n_{1}^{'},n_{2}^{'},\cdot\cdot\cdot,n_{N}^{'})\times\nonumber\\
D(n_{1},n_{2},\cdot\cdot\cdot,n_{N})
\Theta_{x_{1}<x_{2}\cdot\cdot\cdot<x_{N}}e^{2 i k_{so} x_{j}}d\vec{x}.
\label{integral}
\end{eqnarray}
The evaluation of this integral for any $N$ is detailed in Appendix~\ref{appendix_A}.
Once the Hamiltonian matrix has been diagonalized, 
we determine the fully symmetric/anti-symmetric eigenstates by
applying the $N$-particle symmetrizer/anti-symmetrizer.

Since we 
consider 
a fixed particle ordering, we define the sector Hamiltonian
$H_{x_{j_1}<x_{j_2}<\cdot\cdot\cdot<x_{j_N}}$,
\begin{equation}
H_{x_{j_1}<x_{j_2}<\cdot\cdot\cdot<x_{j_N}}=H\Theta_{x_{j_1}<x_{j_2}<\cdot\cdot\cdot<x_{j_N}},
\label{sector hamiltonian}
\end{equation}
which is non-zero only for the particle ordering 
$x_{j_1}<x_{j_2}<\cdot\cdot\cdot<x_{j_N}$.
From the definitions of $H_{x_{j_1}<x_{j_2}<\cdot\cdot\cdot<x_{j_N}}$ and $Y$,
it follows that $YH_{x_{j_1}<x_{j_2}<\cdot\cdot\cdot<x_{j_N}}=H_{x_{j_N}<x_{j_{N-1}}<\cdot\cdot\cdot<x_{j_1}}Y$,
i.e., the $Y$ operator 
does not commute with the sector Hamiltonian.
We can,
however, 
define a sector operator
$Y_{x_{j_1}<x_{j_2}<\cdot\cdot\cdot<x_{j_N}}$
for the  
particle ordering $x_{j_1}<x_{j_2}<\cdot\cdot\cdot<x_{j_N}$, 
which commutes with the sector Hamiltonian 
with the same particle ordering. 
The eigenstates of the $Y$ operator are then obtained by symmetrizing 
or 
anti-symmetrizing 
the eigenstates
of the $Y_{x_{j_1}<x_{j_2}<\cdot\cdot\cdot<x_{j_N}}$ operator.
The key idea is that the particle ordering after
a change of the spatial and spin 
coordinates can be 
``restored'' by exchanging the coordinates of the particles.
Specifically,
$Y_{x_{j_1}<x_{j_2}<\cdot\cdot\cdot<x_{j_N}}$ 
is defined through
\begin{eqnarray}
Y_{x_{j_1}<x_{j_2}<\cdot\cdot\cdot<x_{j_N}}
=(\pm 1)^{[\frac{N}{2}]}\nonumber\\
P_{j_1j_N}P_{j_2j_{N-1}}
\cdot\cdot\cdot P_{j_{[\frac{N}{2}]}j_{N+1-[\frac{N}{2}]}}Y,
\label{sector_Y}
\end{eqnarray}
where $[N/2]$ denotes the integer part of $N/2$, 
and the plus and minus signs apply to identical bosons and fermions, respectively.
It can be proven readily that 
$Y_{x_{j_1}<x_{j_2}<\cdot\cdot\cdot<x_{j_N}}$ 
commutes with 
$H_{x_{j_1}<x_{j_2}<\cdot\cdot\cdot<x_{j_N}}$.
To show that the eigenstates of the $Y$ operator are obtained 
by symmetrizing or anti-symmetrizing the eigenstates of 
$Y_{x_{j_1}<x_{j_2}<\cdot\cdot\cdot<x_{j_N}}$,
let
$\psi_{x_{j_1}<x_{j_2}<\cdot\cdot\cdot<x_{j_N}}$
denote a wave function
in the subspace with the ordering $x_{j_1}<x_{j_2}<\cdot\cdot\cdot<x_{j_N}$ 
with the property
\begin{eqnarray}
Y_{x_{j_1}<x_{j_2}<\cdot\cdot\cdot<x_{j_N}}\psi_{x_{j_1}<x_{j_2}<\cdot\cdot\cdot<x_{j_N}}\nonumber\\
=b \psi_{x_{j_1}<x_{j_2}<\cdot\cdot\cdot<x_{j_N}}.
\label{sector Y eigenstate}
\end{eqnarray}
Taking advantage of
\begin{equation}
\label{modify_symmetrizer}
S^{s} P_{j_1j_N} P_{j_2j_{N-1}}\cdot\cdot\cdot P_{j_{[\frac{N}{2}]}j_{N+1-[\frac{N}{2}]}}=S^{s}, 
\end{equation}
\begin{equation}
S^{a}(-1)^{[\frac{N}{2}]} P_{j_1j_N} P_{j_2j_{N-1}}\cdot\cdot\cdot P_{j_{[\frac{N}{2}]}j_{N+1-[\frac{N}{2}]}}=S^{a},
\label{modify_anti-symmetrizer}
\end{equation}
and the fact that $Y$ commutes with $S^{s(a)}$,
where $S^{s(a)}$ denotes the $N$-particle symmetrizer
(anti-symmetrizer),
we obtain
\begin{eqnarray}
\label{Y_commute}
S^{s(a)} Y_{x_{j_1}<x_{j_2}<\cdot\cdot\cdot<x_{j_N}}\psi_{x_{j_1}<x_{j_2}<\cdot\cdot\cdot<x_{j_N}}=\nonumber\\
Y S^{s(a)}\psi_{x_{j_1}<x_{j_2}<\cdot\cdot\cdot<x_{j_N}}.
\end{eqnarray}
Using Eq.~\eqref{sector Y eigenstate}, Eq.~\eqref{Y_commute} leads to  
$Y S^{s(a)}\psi_{x_{j_1}<x_{j_2}<\cdot\cdot\cdot<x_{j_N}}=bS^{s(a)}\psi_{x_{j_1}<x_{j_2}<\cdot\cdot\cdot<x_{j_N}}$,
i.e., $S^{s(a)}\psi_{x_{j_1}<x_{j_2}<\cdot\cdot\cdot<x_{j_N}}$ is an eigenstate of $Y$ 
with eigenvalue $b$.

To construct the Hamiltonian $H_{eff}$,
we proceed similarly, 
i.e., we also work with a particular particle ordering.
The space $H_L$ is spanned by the states
$D(0,1,\cdot\cdot\cdot,N-1)\Theta_{x_1<x_2<\cdot\cdot\cdot<x_N}|s_1,s_2,\cdot\cdot\cdot,s_N\rangle_y$,
where the  
spin function
can take $2^N$ different arrangements.
Correspondingly, the space $H_H$ is spanned by all other 
unperturbed 
eigenstates 
with the ordering $x_1<x_2<\cdot\cdot\cdot<x_N$.
The Hamiltonian matrix for $H_{eff}$ is constructed and diagonalized,
and states with good symmetry are 
obtained following the same steps as discussed above.

We now discuss the construction of an effective spin Hamiltonian.
Since the particles
can be ordered in $N!$ distinct ways,
the ground state
is $N!$-fold degenerate. 
We integrate out the spatial degrees of freedom of the effective Hamiltonian $H_{eff}$,
yielding a spin Hamiltonian $H_{eff}^{spin}$
that depends only on the spin degrees of freedom.
Specifically, since the Hilbert space $H_L$ contains exactly one spatial wavefunction with ordering
$x_1<x_2<\cdot\cdot\cdot<x_N$, we define $H_{eff}^{spin}$ through
$H_{eff}^{spin}=\int_{-\infty}^{\infty}D^{*}(0,1,\cdot\cdot\cdot,N-1)H_{eff}D(0,1,\cdot\cdot\cdot,N-1)\Theta_{x_1<x_2<\cdot\cdot\cdot<x_N}d\vec{x}$.
Using that the spins are also ordered, $H_{eff}^{spin}$ can be compactly written as
\begin{eqnarray}
H_{eff}^{spin}=\left(E_{0}-\frac{N\hbar^2k_{so}^2}{2m}\right)\hat{I}+\frac{\Omega}{2}\sum_{j=1}^{N}\vec{B}_j\cdot\vec{\sigma}_{j}\nonumber\\
+\frac{\Omega^2}{2\hbar\omega}\sum_{j<k}\vec{\sigma}_{j}^{T}M_{jk}\vec{\sigma}_{k}+\frac{\Omega^2}{4\hbar\omega}\sum_{j}(a_{jj}+b_{jj}),
\label{H_eff_spin}
\end{eqnarray}
where
\begin{equation} 
\vec{\sigma}_j^{T}=(\sigma_{x}^{(j)},\sigma_{z}^{(j)})
\end{equation}
and
\begin{equation}
\vec{B}_j^T=(B_{x}^{(j)},B_{z}^{(j)}).
\end{equation}
$E_{0}$ is the ground state energy of $H_{sr}$, 
$E_{0}=\hbar\omega N^2/2$,
and $M_{jk}$ is a $2\times2$ matrix (see below).
The first, second, and third terms on the right hand side of Eq.~\eqref{H_eff_spin} come from
$H_{eff}^{(0)}, H_{eff}^{(1)},$ and $H_{eff}^{(2)}$, respectively.
The fourth term on the right hand side of Eq.~\eqref{H_eff_spin} also comes from $H_{eff}^{(2)}$;
it accounts for the (spin-independent) onsite interaction,
with $a_{jj}$ and $b_{jj}$ defined below in Eqs.~\eqref{a_jk} and~\eqref{b_jk}.
We find
\begin{equation}
B_{x}^{(j)}=Re(\mathcal{D}_{\substack{0,1,2,\cdot\cdot\cdot,N-1\\0,1,2,\cdot\cdot\cdot,N-1}}^{j})
\end{equation}
and
\begin{equation}
B_{z}^{(j)}=Im(\mathcal{D}_{\substack{0,1,2,\cdot\cdot\cdot,N-1\\0,1,2,\cdot\cdot\cdot,N-1}}^{j}).
\end{equation}
The matrix $M_{jk}$ can be written as
\begin{equation}
M_{jk}=\begin{bmatrix}
a_{jk} &-c_{jk}\\
c_{kj} &b_{jk}
\end{bmatrix},
\end{equation}
where
\begin{equation}
\frac{a_{jk}}{\hbar\omega}=\sum_{\substack{(n_{1},n_{2},n_{3},\cdot\cdot\cdot,n_{N})\\\neq(0,1,2,\cdot\cdot\cdot,N-1)}}\frac{Re(\mathcal{D}_{\substack{0,1,2,\cdot\cdot\cdot,N-1\\n_{1},n_{2},\cdot\cdot\cdot,n_{N}}}^{j})Re(\mathcal{D}_{\substack{0,1,2,\cdot\cdot\cdot,N-1\\n_{1},n_{2},\cdot\cdot\cdot,n_{N}}}^{k})}{E_{0}-E_{n_{1},n_{2},\cdot\cdot\cdot,n_{N}}},
\label{a_jk}
\end{equation}
\begin{equation}
\frac{b_{jk}}{\hbar\omega}=\sum_{\substack{(n_{1},n_{2},n_{3},\cdot\cdot\cdot,n_{N})\\\neq(0,1,2,\cdot\cdot\cdot,N-1)}}\frac{Im(\mathcal{D}_{\substack{0,1,2,\cdot\cdot\cdot,N-1\\n_{1},n_2,\cdot\cdot\cdot,n_N}}^{j})Im(\mathcal{D}_{\substack{0,1,2,\cdot\cdot\cdot,N-1\\n_1,n_2,\cdot\cdot\cdot,n_N}}^{k})}{E_{0}-E_{n_{1},n_{2},\cdot\cdot\cdot,n_{N}}},
\label{b_jk}
\end{equation}
\begin{equation}
\frac{-c_{jk}}{\hbar\omega}=\sum_{\substack{(n_{1},n_{2},n_{3},\cdot\cdot\cdot,n_{N})\\\neq(0,1,2,\cdot\cdot\cdot,N-1)}}\frac{Re(\mathcal{D}_{\substack{0,1,2,\cdot\cdot\cdot,N-1\\n_1,n_2,\cdot\cdot\cdot,n_N}}^{j})Im(\mathcal{D}_{\substack{0,1,2,\cdot\cdot\cdot,N-1\\n_1,n_2,\cdot\cdot\cdot,n_N}}^{k})}{E_{0}-E_{n_{1},n_{2},\cdot\cdot\cdot,n_{N}}},
\label{c_jk}
\end{equation}
and
\begin{equation}
\frac{c_{kj}}{\hbar\omega}=\sum_{\substack{(n_{1},n_{2},n_{3},\cdot\cdot\cdot,n_{N})\\\neq(0,1,2,\cdot\cdot\cdot,N-1)}}\frac{Im(\mathcal{D}_{\substack{0,1,2,\cdot\cdot\cdot,N-1\\n_1,n_2,\cdot\cdot\cdot,n_N}}^{j})Re(\mathcal{D}_{\substack{0,1,2,\cdot\cdot\cdot,N-1\\n_1,n_2,\cdot\cdot\cdot,n_N}}^{k})}{E_{0}-E_{n_{1},n_{2},\cdot\cdot\cdot,n_{N}}}.
\label{c_kj}
\end{equation}

The second and third terms on the right hand side of Eq.~\eqref{H_eff_spin} correspond to 
a single spin term (this is the first-order term) and a spin-spin term, respectively. 
For small $k_{so}$, the first-order term dominates. 
The spin at each slot $j$ follows the effective magnetic $\vec{B_{j}}$ field
and the system has a rotational spin structure~\cite{ho}.
In this regime, the spin correlations are very weak.
The eigenstates of $H_{eff}^{spin}$ 
include equal weights of each spin state,
namely, $P_{|M_s|}$ is proportional to the number of 
spin states that have the same absolute value of $M_s$ (see Table~\ref{table:Table1}).
For example, for $N=3$, two states have $|M_s|=0$ and
six states have $|M_s|=1$, which yields that $P_{|M_s|=0}=1/4$ and $P_{|M_s|=1}=3/4$.

For fairly large $k_{so}$, 
the second-order term dominates over the first-order term, i.e.,
the spin correlations are very strong and the spin structure is non-trivial (see also discussion around Fig.~\ref{intuitive}). 
For large $k_{so}$, we find numerically that the nearest neighbor spin-spin interactions 
dominate.
For example, $|a_{12}|$ is much larger than $|a_{13}|$.
It should be noted, though, that the summands entering into $a_{12}$ and $a_{13}$ are of roughly the same order of magnitude.
Moreover, we find that the nearest-neighbor coefficients $a_{(jk)}, b_{(jk)}, c_{(jk)}$ and $c_{(kj)}$ 
are, for fixed $k_{so}$, approximately equal;
here, the subscript $(jk)$ indicates that $j$ and $k$ are related via $j=k-1$.
We have checked this for $N\le 4$. 
For large $N$, we cannot rule out that the values of the coefficients depend on the slot,
possibly yielding more complicated spin structures than discussed in this work.
For $a_{(jk)}=b_{(jk)}=c_{(jk)}=c_{(kj)}$, the matrix $M_{jk}$ simplifies dramatically, 
\begin{equation}
\label{interaction_matrix}
M_{(jk)}=A\begin{bmatrix}
1 &-1\\
1 &1
\end{bmatrix}
\end{equation}
for $j=k-1$ and $M_{jk}\approx 0$ for $j \neq k-1$.
In Eq.~\eqref{interaction_matrix}, $A$ is a negative dimensionless constant.
In this case, the last term on the right hand side of Eq.~\eqref{H_eff_spin} reduces to
\begin{eqnarray}
\frac{\Omega^2}{2\hbar\omega}\sum_{j<k}\vec{\sigma}_{j}^{T}M_{jk}\vec{\sigma}_{k}=\frac{\Omega^2}{2\hbar\omega}\nonumber\times\\
A\sum_{(jk)}\left[\vec{\sigma}_j\cdot\vec{\sigma}_k
-(\sigma_x^{(j)}\sigma_z^{(k)}-\sigma_{z}^{(j)}\sigma_{x}^{(k)})\right].
\label{second_term_spin_spin_0}
\end{eqnarray}
The first term in square brackets on the right hand side of Eq.~\eqref{second_term_spin_spin_0} corresponds to the usual Heisenberg exchange term. 
The term in round brackets on the right hand side of Eq.~\eqref{second_term_spin_spin_0} is readily indentified 
as the $y$-component of the cross product between two three-dimensional spin vectors.
This term is the one-dimensional analog of the anisotropic Dzyaloshinskii-Moriya exchange term~\cite{Dzyaloshinskii,Moriya,Keffer}.

To obtain the spin structure of this approximate effective spin Hamiltonian, we rewrite Eq.~\eqref{second_term_spin_spin_0} as
\begin{eqnarray}
\label{second_term_spin_spin}
\frac{\Omega^2}{2\hbar\omega}\sum_{j<k}\vec{\sigma}_{j}^{T}M_{jk}\vec{\sigma}_{k}=\frac{\Omega^2}{2\hbar\omega}\nonumber\times\\
A\sum_{(jk)}[2(1-i)\sigma_{-}^{(j)}\sigma_{+}^{(k)}+2(1+i)\sigma_{+}^{(j)}\sigma_{-}^{(k)}],
\end{eqnarray}
where $\sigma_{\pm}^{(j)}=(\sigma_{x}^{(j)}\mp i\sigma_{z}^{(j)})/2$.
If we neglect the first-order term and use the right hand side of Eq.~\eqref{second_term_spin_spin} in Eq.~\eqref{H_eff_spin},
we find that
$H_{eff}^{spin}$ commutes with 
$\sigma_y$.
Moreover, this approximate Hamiltonian $H_{eff}^{spin}$ also has time reversal symmetry, 
i.e., it commutes with $i\sigma_yK$,
where $K$ changes the quantity that it acts on into the complex conjugate of that quantity.
However, $i\sigma_yK$ and $\sigma_y$ do not commute.
Using additionally the property that $K|s\rangle_y=|\bar{s}\rangle_y$,
one can show that the eigenstates with $M_s$ and $-M_s$ of the approximate
effective spin Hamiltonian have the same energy provided $|M_s|>0$, 
i.e., the states with $|M_s|>0$ are two-fold degenerate.  
The first-order term breaks 
the degeneracy of the eigenstates with $M_s$ and $-M_s$.
As a result, the eigenstates of the full Hamiltonian $H_{eff}^{spin}$ (including zero-, first-, and second-order terms) are,
for $k_{so}\to\infty$, approximately superpositions of states that have the same 
absolute value of the $M_{s}$ quantum number.
The $\sigma_{-}^{(j)}\sigma_{+}^{(k)}$ and $\sigma_{+}^{(j)}\sigma_{-}^{(k)}$ terms correspond to nearest neighbor spin hopping.
These terms lead to a lowering of the energy.
The more possibility for the nearest neighbor spin hopping a state has, 
the lower the energy associated with that state is.
For example, the states $|\uparrow\downarrow\rangle_y$ and $|\downarrow\uparrow\rangle_y$ 
are ``connected'' via nearest neighbor hoppings while the states $|\uparrow\uparrow\rangle_y$ and $|\downarrow\downarrow\rangle_y$
are not connected with each other or with $|\uparrow\downarrow\rangle_y$ or $|\downarrow\uparrow\rangle_y$
via nearest neighbor hopping.
Correspondingly, the $N=2$ ground state is a linear combination of the $|\uparrow\downarrow\rangle_y$ and $|\downarrow\uparrow\rangle_y$ states.
For $N=3$, e.g., the state $|\uparrow\uparrow\downarrow\rangle_y$ is connected to $|\uparrow\downarrow\uparrow\rangle_y$
via $\sigma_2^{(-)}\sigma_3^{(+)}$ and to $|\downarrow\uparrow\uparrow\rangle_y$ via $\sigma_1^{(-)}\sigma_3^{(+)}$
while the state $|\downarrow\downarrow\uparrow\rangle_y$ is connected to $|\downarrow\uparrow\downarrow\rangle_y$
via $\sigma_2^{(+)}\sigma_3^{(-)}$ and to $|\uparrow\downarrow\downarrow\rangle_y$ via $\sigma_1^{(+)}\sigma_3^{(-)}$.
Correspondingly, the $N=3$ ground state is a linear combination of all $|M_s|=1$ states. 
Table~\ref{table:Table1} shows the values of $P_{|M_s|=m}$ in the limits $k_{so}\to 0$ and $k_{so}\to\infty$ for $N=2-4$. 
\begin{table}
\begin{center}
\begin{tabular}{c|c|c}
\hline
$N$ & $k_{so}\to 0$ &    $k_{so}\to \infty$\\
\hline
$2$ & $P_0=1/2, P_2=1/2$ &    $P_0=1, P_2=0$\\
$3$ & $P_1=3/4, P_3=1/4$ &    $P_1=1, P_3=0$\\
$4$ & $P_0=3/8, P_2=1/2, P_4=1/8$ &    $P_0=1, P_2=0, P_4=0$\\
\hline
\end{tabular}
\caption{Spin correlations for the ground state in the limiting cases $k_{so}\rightarrow 0$ and $k_{so}\rightarrow\infty$ for various $N$.
The probability $P_{|M_s|}$ that the ground state has 
the absolute value of $M_s$ is reported.
For small $k_{so}$, all spin states are equally weighted, which means that $P_{|M_s|}$ is proportional to 
the number of spin states that have the same $|M_s|$. For large $k_{so}$,
the spin states are not equally weighted. 
In this case, the ground state contains only spin states with the minimum allowed $|M_s|$.}
\label{table:Table1}
\end{center}
\end{table}

\begin{figure}
\vspace*{-0.5cm}
\hspace*{-0.5cm}
\includegraphics[width=0.32\textwidth,trim=0.05cm 0cm 0.05cm 0cm, clip=true]{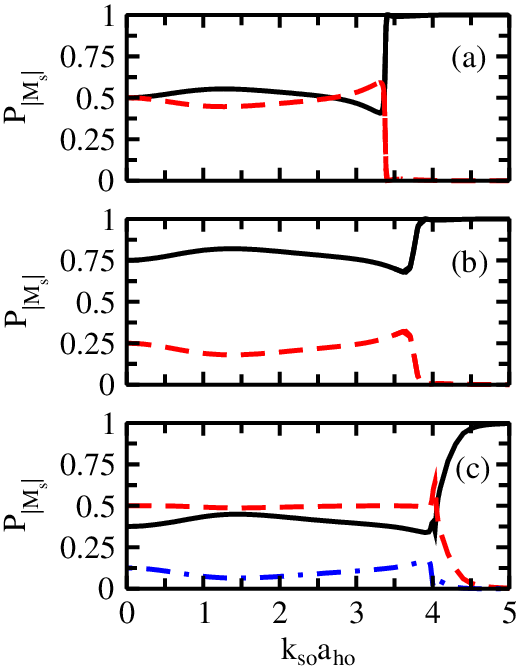}
\vspace*{0cm}
\caption{Expectation value of $P_{|M_s|=m}$ for (a) $N=2$, (b) $N=3$, and (c) $N=4$ with $g=\infty$
and $\Omega=\hbar\omega/2$ as a function of $k_{so}$.
(a) The solid and dashed lines show $P_{|M_{s}|=0}$ and $P_{|M_{s}|=2}$, respectively. 
(b) The solid and dashed lines show $P_{|M_{s}|=1}$ and $P_{|M_{s}|=3}$, respectively.
(c) The solid, dashed, and dot-dashed lines show $P_{|M_s|=0}, P_{|M_s|=2}$ and $P_{|M_s|=4}$, respectively.}
\label{correlation}
\end{figure}

Figure~\ref{correlation} shows the probability to find the system in a state with a given $|M_{s}|$ as a function of $k_{so}$ for the two-, 
three-, and four-particle systems with infinitely large $g$ (see the next subsection for the calculational details). 
For large $k_{so}$, $P_{|M_{s}|=0}=1$ for the two-particle system, 
$P_{|M_{s}|=1}=1$ for the three-particle system, and $P_{|M_{s}|=0}=1$ for the four-particle system.

\subsection{Application to systems with $N=2-4$}
\label{subsection_N2-4}
This section evaluates the expectation values of the 
spin operators defined in Eqs.~\eqref{S_x}-\eqref{P_Ms} 
for the
three- and four-particle systems with infinite $g$.
For $N=2$, the expectation values of $S_x(x)$ and $S_z(x)$ for infinitely large $g$ have been calculated in 
Sec.~\ref{sec_two_particle}.
For $N=3$, the non-symmetrized ground state 
wave function
of the effective Hamiltonian $H_{eff}$ is
\begin{eqnarray}
\label{eigenstate_3particle}
\psi_{gr}=\frac{D(0,1,2)}{\sqrt{2}}\Theta_{x_1<x_2<x_3}\nonumber\times\\
\Big[C_{1}(|\uparrow\uparrow\uparrow\rangle_y-|\downarrow\downarrow\downarrow\rangle_y)
+C_{2}(|\uparrow\uparrow\downarrow\rangle_y-|\uparrow\downarrow\downarrow\rangle_y)\nonumber\\
+C_{3}(|\uparrow\downarrow\uparrow\rangle_y-|\downarrow\uparrow\downarrow\rangle_y)
+C_{4}(|\downarrow\uparrow\uparrow\rangle_y-|\downarrow\downarrow\uparrow\rangle_y)\Big].
\end{eqnarray}
The coefficients $C_1$-$C_4$ are obtained by diagonalizing the effective low-energy Hamiltonian $H_{eff}$.
Using this wave function, the expectation values of $S_{x}(x)$ and $S_{z}(x)$ for the three-particle system are
\begin{equation}
\label{Sx_3particle}
\langle S_{x}(x)\rangle=\frac{\hbar}{2}[C_{1x}n_{1x}(x)+C_{2x}n_{2x}(x)]
\end{equation}
and
\begin{equation}
\label{Sz_3particle}
\langle S_{z}(x)\rangle=\frac{\hbar}{2}C_{z}n_{z}(x).
\end{equation}
The coefficients $C_{1x}, C_{2x}$, and $C_{z}$,
which depend on
the coefficients $C_1-C_{4}$ 
in Eq.~\eqref{eigenstate_3particle}, and the expressions 
for the density functions
$n_{1x}(x)$, $n_{2x}(x)$, and $n_{z}(x)$ are given in Appendix~\ref{appendix_B}.
Figure~\ref{3particle} shows the 
spin coefficients and spin densities for $N=3$ 
with infinite $g$ for different $k_{so}$.
For fixed $x$,
$\langle S_{x}(x)\rangle$ and $\langle S_{z}(x)\rangle$ 
oscillate with $k_{so}$ for $k_{so}\le k_{so}^{cr}$, where  
$a_{ho}k_{so}^{cr}\approx 3.7$.
The oscillations disappear for $k_{so}$
greater than $k_{so}^{cr}$.
\begin{figure}
\vspace*{0cm}
\hspace*{-0.5cm}
\includegraphics[width=0.43\textwidth]{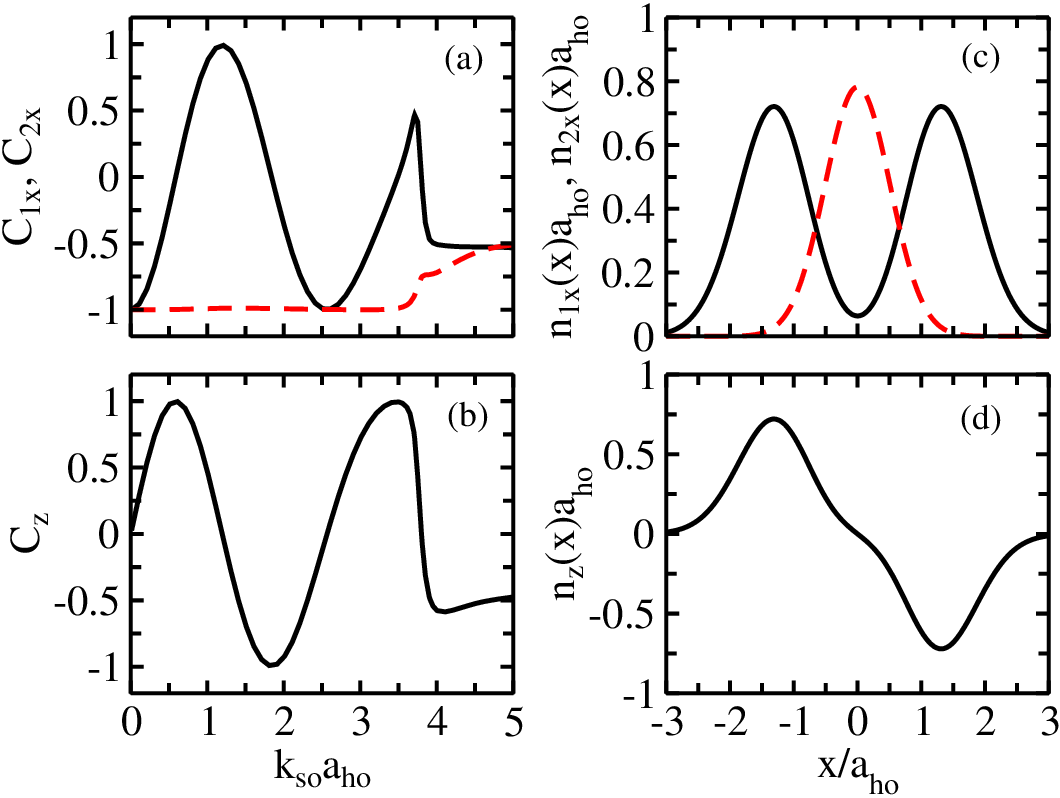}
\vspace*{0cm}
\caption{Spin structure for the three-particle system with $g=\infty$ and $\Omega=\hbar\omega/2$ as a function of $k_{so}$. 
(a) The solid and dashed lines show the spin structure coefficients $C_{1x}$ and $C_{2x}$, respectively. 
(b) The solid and dashed lines show the spin structure coefficient $C_{z}$. 
(c) The solid and dashed lines show the spin structure densities $n_{1x}(x)$ and $n_{2x}(x)$, respectively.
(d) The solid line shows the spin structure density $n_{z}(x)$.}
\label{3particle}
\end{figure}

For $N=4$,
the non-symmetrized ground state 
wave function
of the effective Hamiltonian $H_{eff}$ is
\begin{eqnarray}
\label{eigenstate_4particle}
\psi_{gr}=D(0,1,2,3)
\Bigg\{\frac{1}{\sqrt{2}}\Big[C_{1}\left(|\uparrow\uparrow\uparrow\uparrow\rangle_y+|\downarrow\downarrow\downarrow\downarrow\rangle_y\right)\nonumber\\
+C_{2}\left(|\uparrow\uparrow\uparrow\downarrow\rangle_y+|\uparrow\downarrow\downarrow\downarrow\rangle_y\right)
+C_{3}\left(|\uparrow\uparrow\downarrow\uparrow\rangle_y+|\downarrow\uparrow\downarrow\downarrow\rangle_y\right)\nonumber\\
+C_{4}\left(|\uparrow\downarrow\uparrow\uparrow\rangle_y+|\downarrow\downarrow\uparrow\downarrow\rangle_y\right)
+C_{5}\left(|\downarrow\uparrow\uparrow\uparrow\rangle_y+|\downarrow\downarrow\downarrow\uparrow\rangle_y\right)\nonumber\\
+C_{8}\left(|\uparrow\downarrow\downarrow\uparrow\rangle_y+|\downarrow\uparrow\uparrow\downarrow\rangle_y\right)\Big]
+C_6|\uparrow\uparrow\downarrow\downarrow\rangle_y+C_7|\uparrow\downarrow\uparrow\downarrow\rangle_y\nonumber\\
+C_9|\downarrow\downarrow\uparrow\uparrow\rangle_y
+C_{10}|\downarrow\uparrow\downarrow\uparrow\rangle_y\Bigg\}\Theta_{x_1<x_2<x_3<x_4}.
\end{eqnarray}
The coefficients $C_1$-$C_{10}$ are obtained by diagonalizing the effective low-energy Hamiltonian $H_{eff}$.
Using this wave function, the expectation values of $S_{x}(x)$ and $S_{z}(x)$ for the four-particle system are
\begin{equation}
\label{Sx_4particle}
\langle S_{x}(x)\rangle=\frac{\hbar}{2}[C_{1x}n_{1x}(x)+C_{2x}n_{2x}(x)]
\end{equation}
and
\begin{equation}
\label{Sz_4particle}
\langle S_{z}(x)\rangle=\frac{\hbar}{2}[C_{1z}n_{1z}(x)+C_{2z}n_{2z}(x)].
\end{equation}
Expressions for
the coefficients $C_{1x}, C_{2x}, C_{1z},$ and $C_{2z}$ 
and the density functions
$n_{1x}(x), n_{2x}(x), n_{1z}(x),$ and $n_{2z}(x)$ are given in Appendix~\ref{appendix_B}.
Figures~\ref{4particle}(a) and~\ref{4particle}(b) show that the spin structure 
coefficients go to approximately zero at $a_{ho}k_{so}^{cr}\approx 4.3$. 
\begin{figure}
\vspace*{0cm}
\hspace*{-0.5cm}
\includegraphics[width=0.43\textwidth]{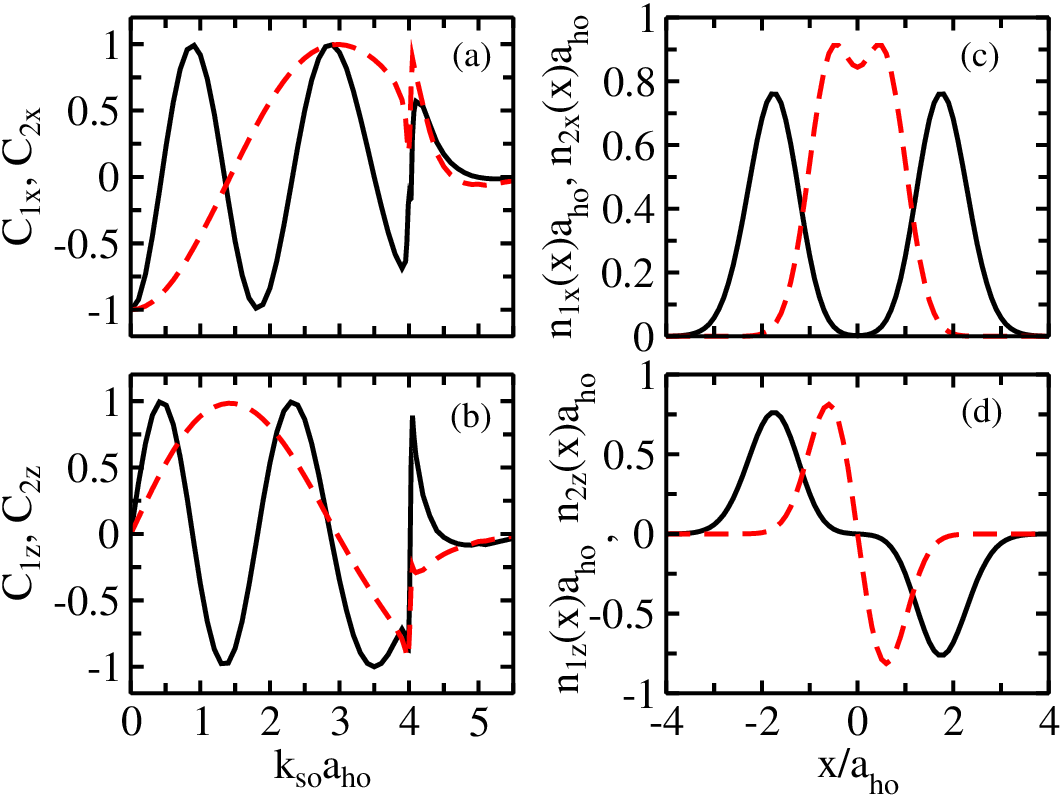}
\vspace*{0cm}
\caption{Spin structure for the four-particle system with $g=\infty$ and $\Omega=\hbar\omega/2$ as a function of $k_{so}$. 
(a) The solid and dashed lines show the spin structure coefficients $C_{1x}$ and $C_{2x}$.
(b) The solid and dashed lines show the spin structure coefficients $C_{1z}$ and $C_{2z}$. 
(c) The solid and dashed lines show the spin structure densities $n_{1x}(x)$ and $n_{2x}(x)$.
(d) The solid and dashed lines show the spin structure densities $n_{1z}(x)$ and $n_{2z}(x)$.}
\label{4particle}
\end{figure}

\begin{figure}
\vspace*{0cm}
\hspace*{-0.1cm}
\includegraphics[width=0.3\textwidth, trim=0.05cm 0cm 0.05cm 0cm, clip=true]{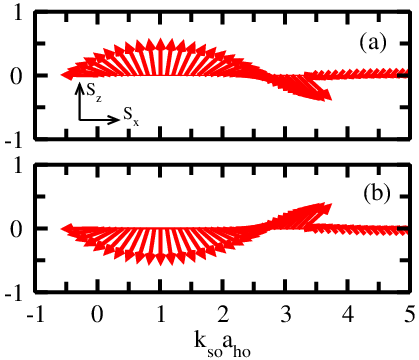}
\caption{The spin structure for two identical particles with $g=\infty$ and $\Omega=\hbar\omega/2$.
Panel (a) shows the spin vector at the leftmost slot while 
panel (b) shows the spin vector at the rightmost slot.
For $k_{so}\lesssim k_{so}^{cr}\approx 3.4/a_{ho}$, the spin vector at each slot follows the local effective $\vec{B}$ field.
The spin vector rotates with increasing $k_{so}$.
For $k_{so}\gtrsim k_{so}^{cr}$, the spin-spin interaction is dominant, resulting in an approximately vanishing spin vector at each slot.}
\label{2particlearrow}
\end{figure}

\begin{figure}
\hspace*{0cm}
\vspace*{0cm}
\includegraphics[width=0.3\textwidth]{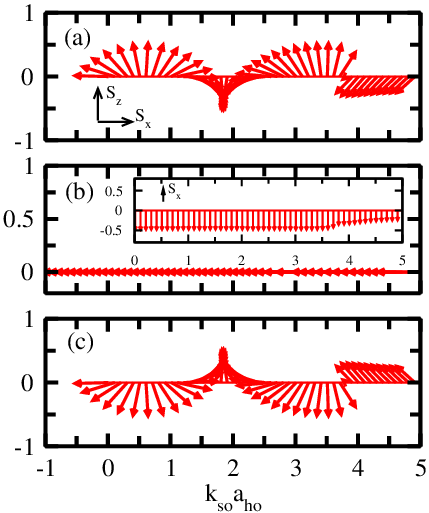}
\caption{The spin structure for three identical particles with $g=\infty$ and $\Omega=\hbar\omega/2$.
Panel (a) shows the spin vector at the leftmost slot, 
panel (b) shows the spin vector at the middle slot, and
panel (c) shows the spin vector at the rightmost slot.
For $k_{so}\lesssim k_{so}^{cr}\approx 3.7/a_{ho}$, the spin vector at each slot follows the local effective $\vec{B}$ field.
The spin vector rotates with increasing $k_{so}$.
For $k_{so}\gtrsim k_{so}^{cr}$, the spin-spin interaction is dominant, resulting in a spin vector 
with approximately constant magnitude and orientation at each slot.
The inset in (b) replots the spin vector using a different orientation of the coordinate system.  
}
\label{3particlearrow}
\end{figure}

\begin{figure}
\vspace*{0cm}
\hspace*{0cm}
\includegraphics[width=0.3\textwidth, trim=0.05cm 0cm 0.05cm 0cm, clip=true]{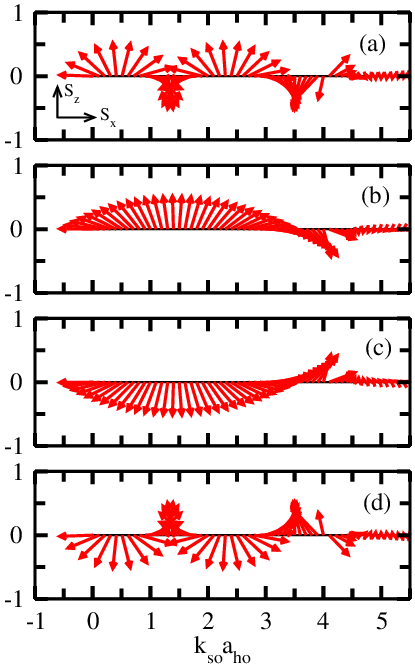}
\caption{The spin structure for four identical particles with $g=\infty$ and $\Omega=\hbar\omega/2$.
Panel (a) shows the spin vector at the leftmost slot.
Panel (b) shows the spin vector at the second slot from the left.
Panel (c) shows the spin vector at the third slot from the left.
Panel (d) shows the spin vector at the rightmost slot.
For $k_{so}\lesssim k_{so}^{cr}\approx 4.3/a_{ho}$, the spin vector at each slot follows the local effective $\vec{B}$ field.
The spin vector rotates with increasing $k_{so}$.
For $k_{so}\gtrsim k_{so}^{cr}$, the spin-spin interaction is dominant, resulting in an approximately vanishing spin vector at each slot.}
\label{4particlearrow}
\end{figure}

To visualize the spin structure,
Figs.~\ref{2particlearrow}-\ref{4particlearrow}
show the spin vector $(\langle S_x(x) \rangle, \langle S_z(x) \rangle)$ 
at each ``slot'' as a function of $k_{so}$ for $N=2-4$ 
with $g=\infty$. 
In each figure, the top to bottom panels
correspond to the  
leftmost to the rightmost slot.
For $k_{so}\lesssim k_{so}^{cr}$,
the spin vector rotates with increasing $k_{so}$. For $k_{so}\gtrsim k_{so}^{cr}$,
the spin-spin interaction term dominates.
For $N=2$ and $N=4$, the magnitude of the spin vector at each slot is approximately zero for $k_{so}\gtrsim k_{so}^{cr}$.
For $N=3$, in contrast,
the magnitude of the spin vector at each slot is fixed and the orientation of the spin vector is,
to a good approximation, independent of $k_{so}$. 
Taking the ground state of the approximate effective Hamiltonian to calculate 
$\langle S_{x}(x)\rangle$ and $\langle S_{z}(x)\rangle$,
we find $\langle S_{x}(x)\rangle\approx\langle S_{z}(x)\rangle\approx -\hbar/2$ for the leftmost slot
and $\langle S_{x}(x)\rangle\approx -\langle S_{z}(x)\rangle\approx -\hbar /2$ for the rightmost slot for $k_{so}\gtrsim k_{so}^{cr}$.
This behavior is a signature of the spin-spin correlations.
For even $N$ systems, the ground state in the large $k_{so}$ limit is a superposition of spin states with $|M_s|=0$.
Two states $|s_1,s_2,\cdot\cdot\cdot,s_N\rangle_y$ 
and $|s_1^{'},s_2^{'},\cdot\cdot\cdot,s_N^{'}\rangle_y$, both with $M_s=0$,
contain an even number of $s_j$ and $s_j^{'}$ for which $s_j\neq s_j^{'}$. 
For $N=2$, e.g., to get the state $|\uparrow\downarrow\rangle_y$  
from the state $|\downarrow\uparrow\rangle_y$, two spin flips are needed.
For odd $N$ systems, the ground state in the large $k_{so}$ limit is a superposition of spin states with $|M_s|=1$.
Two states $|s_1,s_2,\cdot\cdot\cdot,s_N\rangle_y$ and $|s_1^{'},s_2^{'},\cdot\cdot\cdot,s_N^{'}\rangle_y$, both with $|M_s|=1$,
contain an odd number of $s_j$ and $s_j^{'}$ for which $s_j\neq s_j^{'}$.
For $N=3$, e.g., to get the state $|\uparrow\uparrow\downarrow\rangle_y$ from 
the state $|\uparrow\downarrow\downarrow\rangle_y$, one spin flip is needed.
The non-vanishing $\langle S_x(x)\rangle$ and $\langle S_z(x)\rangle$ arise from a superposition of states 
that contain $N-1$ identical spins ($s_j=s_j^{'}$) and one pair of opposite spins $(s_j\neq s_j^{'})$.
As a result, the spin vector $(\langle S_x(x)\rangle ,\langle S_z(x)\rangle)$ vanishes for even $N$ systems with large $k_{so}$
while that for odd $N$ systems is finite and approximately constant. 

\section{conclusion}
\label{sec_conclusion}

Spin Hamiltonian play an important role in understanding material properties such as
the transition from ferromagnetic to anti-ferromagnetic order.
Much effort has gone into identifying clean model systems with which to emulate spin dynamics.
Notable examples include trapped ion systems, 
ranging from small two- or three-ion chains~\cite{ion0,ion1} to large two-dimensional ion 
crystals~\cite{ion2,ion3}, and ultracold atoms~\cite{mean field2, mean field3, H pu,artem,Levinsen,Deuretzbacher,ultra atoms,ultra atoms 1}.
This paper considered ultracold harmonically trapped one-dimensional atoms with infinitely 
large two-body contact interactions subject to spin-orbit and Raman couplings and derived 
an effective low-energy spin Hamiltonian that is accurate to second order in the Raman coupling strength $\Omega$. 
It was shown explicitly for $N=2-4$ particles that tuning the spin-orbit coupling strength from a regime where
the spins independently follow the effective external magnetic field generated by the Raman coupling
to a regime where the spin dynamics is governed by the nearest neighbor spin-spin interactions is feasible.
While the examples presented are for small particle numbers,
Appendix~\ref{appendix_A} derived a number of identities applicable to systems of arbitrary size
that should facilitate extensions to larger $N$ and that may find applications in calculating 
the momentum distribution or related observables for strongly-interacting one-dimensional 
atomic gases without spin-orbit coupling. 

The relevance of the effective spin Hamiltonian derived in this paper is two-fold.
(i) As already alluded to above, this paper introduced a new means to realize a tunable effective spin Hamiltonian.
(ii) The spin Hamiltonian language provides a physically transparent means to understand the intricate dynamics 
of strongly correlated spin-orbit coupled ultracold atomic systems.

Throughout this paper, explicit calculations were performed for the Raman coupling strength $\Omega=\hbar\omega/2$.
For this coupling strength, the energy difference for $k_{so}\ge k_{so}^{cr}$ between the ground state
and the first excited state is of the order of $10^{-4}\hbar\omega$ for $N=2-4$ particles.
This small energy difference may make it challenging to experimentally occupy the ground state.
Since the second-order term is proportional to $\Omega^2/(\hbar\omega)$,
the critical $k_{so}^{cr}$ decreases with increasing $\Omega$. 
For $\Omega=4\hbar\omega$ and $N=3$, e.g., we find $k_{so}^{cr}a_{ho}\approx 3.5$ and an energy
splitting between the ground state and the first excited state with the same symmetry of the order of $10^{-2}\hbar\omega$.
This energy splitting appears much more tractable experimentally.

An important question is then whether the effective second-order low-energy spin Hamiltonian 
is applicable for such large $\Omega$. 
To investigate this question, we compared the results obtained by solving the second-order effective
Hamiltonian with the results obtained from the diagonalization of the full Hamiltonian.
The agreement is very good for $k_{so}\geq k_{so}^{cr}$ (we excluded the regime $k_{so}\gg k_{so}^{cr}$), 
suggesting that the third-order term, which is proportional to $\Omega^3/(\hbar\omega)^2$,
is suppressed. 
Naively, one expects the third-order term to be enhanced by $\Omega/(\hbar\omega)$ compared to 
the second-order term  for $\Omega>\hbar\omega$, making a perturbative approach meaningless.
However, the full perturbative expression also contains the matrix elements and energy denominator.
It turns out that the product of three matrix elements is highly suppressed compared to the product of two matrix elements.
We conclude that the second-order effective spin Hamiltonian capture the physics, 
including the spin structure up to, at first sight, surprisingly large $\Omega/(\hbar\omega)$
for relatively large $k_{so}$.

A key ingredient that went into deriving the effective low-energy spin Hamiltonian is that 
the particles in one-dimensional space can be ordered if the two-body coupling constant $g$ is infinitely large. 
For finite $g$, this is not the case, i.e., particles are allowed to pass through each other.
In this case, a low-energy Hamiltonian that depends on the spatial and spin degrees of freedom was derived 
(in fact, the effective spin Hamiltonian for $g=\infty$ was derived by taking this Hamiltonian and integrating out the spatial degrees of freedom).
The low-energy Hamiltonian was tested for two particles and shown to reproduce the full Hamiltonian dynamics well.
Moreover, it was shown to provide a powerful theoretical framework within which to interpret the full Hamiltonian results.
We believe that the formalism can be applied to larger one-dimensional system and extended to higher-dimensional systems. 

\section{Acknowledgement}
Support by the National Science Foundation through grant number PHY-1205443 
and 
discussions with X. Y. Yin,
S. E. Gharashi
and T.-L. Ho 
are gratefully acknowledged. 

\appendix
\begin{widetext}
\section{Calculation of involved integrals}
\label{appendix_A}

This section contains the evaluation of the integral given in Eq.~\eqref{integral}. 
We denote the $n$th harmonic 
oscillator
eigenstate by $\varphi_{n}(x)$,
$\varphi_{n}(x)=N_n H_{n}(x/a_{ho})e^{-x^2/(2a_{ho}^{2})}$,
where $N_n=1/\sqrt{\sqrt{\pi}n! 2^n a_{ho}}$ 
is the normalization constant
and $H_{n}(x)$ the $n$th Hermite polynomial.
Throughout this appendix, 
we set $a_{ho}=1$, i.e., we work with dimensionless spatial coordinates.
Expanding the Slater determinant $D(n_1, n_2,\cdot\cdot\cdot,n_N)$,
we have
\begin{equation}
\label{D_defination}
D(n_1,n_2,\cdot\cdot\cdot,n_N)=\sum_{p_1,p_2,\cdot\cdot\cdot,p_N}(-1)^{P_{p_1,p_2,\cdot\cdot\cdot,p_N}}\Pi_{l=1}^{N}\varphi_{n_{p_{l}}}(x_l),
\end{equation}
where $p_1,p_2,\cdot\cdot\cdot,p_N$ denotes 
a permutation of $1,2,\cdot\cdot\cdot,N$ and $P_{p_1,p_2,\cdot\cdot\cdot,p_N}$ the number 
of permutations needed to obtain the order $p_1, p_2,\cdot\cdot\cdot,p_N$ from the ordinary order $1,2,\cdot\cdot\cdot,N$.
The sum in Eq.~\eqref{D_defination} contains $N!$ terms.
Note that since the eigenstates $\phi_{n_1,n_2,\cdot\cdot\cdot,n_N}(\vec{x})$ in Eq.~\eqref{wavefunc_infg} are only non-zero 
for a particular 
particle ordering,
we do not need 
a prefactor of
$1/\sqrt{N!}$ in 
front of the Slater determinant to normalize the eigenstates.   
Equation~\eqref{integral} 
contains two different Slater determinants 
($D$ functions),
one with arguments $n_1,\cdot\cdot\cdot,n_N$ and the other
with arguments $n_1^{'},\cdot\cdot\cdot,n_n^{'}$.
To simplify the notation,
we use $m_1,\cdot\cdot\cdot,m_N$ instead of $n_1^{'},\cdot\cdot\cdot,n_n^{'}$ in what follows.
The corresponding permutations are denoted by $q_1,\cdot\cdot\cdot,q_N$.
With these conventions, 
Eq.~\eqref{integral} 
becomes
\begin{multline}
\label{double_sum0}
\mathcal{D}_{\substack{n_{1},n_{2},\cdot\cdot\cdot,n_{N}\\m_{1},m_{2},\cdot\cdot\cdot,m_{N}}}^{j}=\int_{-\infty}^{\infty}\left(\sum_{p_1,p_2,\cdot\cdot\cdot,p_N}(-1)^{P_{p_1,p_2,\cdot\cdot\cdot,p_N}}\Pi_{l=1}^{N}\varphi_{n_{p_{l}}}(x_l)\right)\\
\left(\sum_{q_1,q_2,\cdot\cdot\cdot,q_N}(-1)^{P_{q_1,q_2,\cdot\cdot\cdot,q_N}}\Pi_{l=1}^{N}\varphi_{m_{q_l}}(x_{l})\right)
\Theta_{x_1<x_2<\cdot\cdot\cdot <x_N}e^{2ik_{so} x_{j}}d\vec{x}.
\end{multline}
Since the integral in Eq.~\eqref{double_sum0} can be interpreted as a Fourier transform with 
respect to the coordinate $x_{j}$, we first evaluate the integral over the coordinates $x_1,\cdot\cdot\cdot,x_{j-1}, x_{j+1},\cdot\cdot\cdot,x_N$.
Except for the sector function
$\Theta_{x_1<x_2<\cdot\cdot\cdot<x_N}$, the  
integrand in Eq.~\eqref{double_sum0} is symmetric under the exchange 
of any two variables that are smaller and larger than $x_j$.
By changing the order of the integration
variables that are smaller and larger than $x_{j}$,
we get~\cite{H pu,Deuretzbacher,girardeau1,girardeau2}
\begin{equation}
\label{double_sum}
\mathcal{D}_{\substack{n_{1},n_{2},\cdot\cdot\cdot,n_{N}\\m_{1},m_{2},\cdot\cdot\cdot,m_{N}}}^{j}=\sum_{p_1,p_2,\cdot\cdot\cdot,p_N}\sum_{q_1,q_2,\cdot\cdot\cdot,q_N}(-1)^{P_{p_1,p_2,\cdot\cdot\cdot,p_N}}(-1)^{P_{q_1,q_2,\cdot\cdot\cdot,q_N}} I_{\substack{n_{p_1},n_{p_2},\cdot\cdot\cdot,n_{p_N}\\m_{q_1},m_{q_2},\cdot\cdot\cdot,m_{q_N}}}^{j},
\end{equation}
where
\begin{multline}
I_{\substack{n_{p_1},n_{p_2},\cdot\cdot\cdot,n_{p_N}\\m_{q_1},m_{q_2},\cdot\cdot\cdot,m_{q_N}}}^{j}=\frac{1}{(j-1)!(N-j)!}
\int_{-\infty}^{\infty}dx_{j}\Bigg[\varphi_{n_{p_j}}(x_j)\varphi_{m_{q_j}}(x_{j})e^{2ik_{so} x_{j}}\times\\
\left(\Pi_{k<j}I^{(1)}_{n_{p_k},m_{q_k}}(x_j)\right)
\left(\Pi_{l>j}I^{(2)}_{n_{p_l},m_{q_l}}(x_j)\right)\Bigg]
\label{integral_I}
\end{multline}
with
\begin{equation}
I^{(1)}_{n_{p_k},m_{q_k}}(x_j)=\int_{-\infty}^{x_j}
\varphi_{n_{p_k}}(x)\varphi_{m_{q_k}}(x)dx
\label{Integral1}
\end{equation} 
and 
\begin{equation}
\label{Integral2}
I^{(2)}_{n_{p_l},m_{q_l}}(x_j)=\int_{x_j}^{\infty}
\varphi_{n_{p_l}}(x)\varphi_{m_{q_l}}(x)dx.
\end{equation}
According to the generalized Feldheim identity, the product of any number of Hermite polynomials can be expanded into
a finite sum of Hermite polynomials~\cite{math book},
\begin{equation}
\label{Hermite_expansion}
H_{N_1}(x)\cdot\cdot\cdot H_{N_m}(x)=\sum_{\nu_1\cdot\cdot\cdot\nu_{m-1}}a_{\nu_1\cdot\cdot\cdot\nu_{m-1}}H_{M}(x),
\end{equation}  
where
\begin{equation}
\label{M}
M=\sum_{l=1}^{m-1}(N_{l}-2\nu_{l})+N_{m}
\end{equation}
and
\begin{equation}
\label{a_coefficient}
a_{\nu_1\cdot\cdot\cdot\nu_{m-1}}=\Pi_{l=1}^{m-1}\binom{N_{l+1}}{\nu_{l}}\binom{\sum_{k=1}^{l-1}(N_{k}-2\nu_{k}+N_{l})}{\nu_{l}}2^{\nu_l}\nu_{l}!.
\end{equation}
The limits of the summation indices $\nu_{k}$ in Eq.~\eqref{Hermite_expansion} are given by
\begin{multline}
\label{limit}
0\le\nu_1\le\min\left(N_1,N_2\right),\ 
0\le\nu_2\le\min\left(N_{3},N_1+N_2-2\nu_1\right)\ ,\cdot\cdot\cdot,\ 
0\le\nu_{m-1}\le\min\left(N_m, \sum_{k=1}^{m-2}(N_k-2\nu_k)+N_{m-1}\right).
\end{multline}
Using Eqs.~\eqref{Hermite_expansion}-\eqref{limit}
to rewrite the product of the two Hermite polynomials 
contained in 
$I^{(1)}(x_j)$ and $I^{(2)}(x_j)$,
we obtain
\begin{equation}
I^{(1)}_{n_{p_k},m_{q_k}}(x_{j})=N_{n_{p_k}}N_{m_{q_k}}\sum_{\nu_1^{(k)}=0}^{\min(n_{p_k},m_{q_k})}a_{\nu_1^{(k)}}\int_{-\infty}^{x_j}H_{n_{p_k}+m_{q_k}-2\nu_1^{(k)}}(x)\exp{(-x^2)}dx
\label{Integral11}
\end{equation}
and
\begin{equation}
I^{(2)}_{n_{p_l},m_{q_l}}(x_{j})=N_{n_{p_l}}N_{m_{q_l}}
\sum_{\nu_1^{(l)}=0}^{\min(n_{p_l},m_{q_l})}
a_{\nu_1^{(l)}}
\int_{x_j}^{\infty}
H_{n_{p_l}+m_{q_l}-2\nu_1^{(l)}}(x)\exp{(-x^2)}dx.
\label{Integral22}
\end{equation}
Instead of working with Eqs.~\eqref{Integral11} and \eqref{Integral22} directly, we replace
the Hermite polynomials in the integrals using the generating function $g(x,t)$ of the Hermite polynomials,
\begin{equation}
g(x,t)=e^{-t^2+2tx}=\sum_{n=0}^{\infty}H_{n}(x)\frac{t^n}{n!}.
\label{generating_function}
\end{equation} 
From Eq.~\eqref{generating_function}, we have
\begin{equation}
\label{Hermite_from_generating_function}
H_{n}(x)=\frac{\partial^n g(x,t)}{\partial t^n}\Bigg|_{t=0}.
\end{equation}
Inserting Eq.~\eqref{Hermite_from_generating_function} into Eqs.~\eqref{Integral11}-\eqref{Integral22},
we find
\begin{equation}
I^{(1)}_{n_{p_k},m_{q_{k}}}(x_{j})=N_{n_{p_k}}N_{m_{q_k}}\sum_{\nu_1^{(k)}=0}^{\min(n_{p_k},m_{q_k})}a_{\nu_1^{(k)}}\frac{\partial^{n_{p_k}+m_{q_k}-2\nu_1^{(k)}}}{\partial t^{n_{p_k}+m_{q_k}-2\nu_1^{(k)}}}
\left[\int_{-\infty}^{x_j}\exp(-t^2+2 t x-x^2)dx\right]
\Bigg|_{t=0}
\label{generating_Integral11}
\end{equation}
and
\begin{equation}
I^{(2)}_{n_{p_l},m_{q_l}}(x_{j})=N_{n_{p_l}}N_{m_{q_l}}\sum_{\nu_1^{(l)}=0}^{\min(n_{p_l},m_{q_l})}a_{\nu_1^{(l)}}\frac{\partial^{n_{p_l}+m_{q_l}-2\nu_1^{(l)}}}{\partial t^{n_{p_l}+m_{q_l}-2\nu_1^{(l)}}}
\left[\int_{x_j}^{\infty}\exp(-t^2+2 t x-x^2)dx\right]
\Bigg|_{t=0}.
\label{generating_Integral22}
\end{equation}
The key point is that the Gaussian integrals
can be calculated analytically. This yields
\begin{multline}
I^{(1)}_{n_{p_k},m_{q_k}}(x_{j})=N_{n_{p_k}}N_{m_{q_k}}\sum_{\nu_1^{(k)}=0}^{\min(n_{p_k},m_{q_k})}a_{\nu_1^{(k)}}\Bigg[\left(\frac{\sqrt{\pi}}{2}-\frac{\sqrt{\pi}}{2}\text{erf}(x_j)\right)\delta_{n_{p_k}+m_{q_k}-2\nu_1^{(k)},0}\\
+\exp(-x_{j}^2)H_{n_{p_k}+m_{q_k}-2\nu_1^{(k)}-1}(x_{j})(1-\delta_{n_{p_k}+m_{q_k}-2\nu_1^{(k)},0})\Bigg]
\label{Integral1_f}
\end{multline}
and
\begin{multline}
\label{Integral2_f}
I^{(2)}_{n_{p_l},m_{q_l}}(x_{j})=N_{n_{p_l}}N_{m_{q_l}}\sum_{\nu_1^{(l)}=0}^{\min(n_{p_l},m_{q_l})}a_{\nu_1^{(l)}}\Bigg[\left(\frac{\sqrt{\pi}}{2}+\frac{\sqrt{\pi}}{2}\text{erf}(x_j)\right)\delta_{n_{p_l}+m_{q_l}-2\nu_1^{(l)},0}\\
-\exp(-x_{j}^2)H_{n_{p_l}+m_{q_l}-2\nu_1^{(l)}-1}(x_{j})(1-\delta_{n_{p_l}+m_{q_l}-2\nu_1^{(l)},0})\Bigg],
\end{multline}
where $\text{erf}(x)$ is 
the error function,
\begin{equation}
\text{erf}(x)=\frac{2}{\sqrt{\pi}}\int_{0}^{x}
\exp(-t^2)
dt,
\end{equation}
and $\delta_{n,m}$ the Kronecker delta function.
Plugging Eqs.~\eqref{Integral1_f} and \eqref{Integral2_f} into Eq.~\eqref{integral_I} and rearranging the order of the sums and products, 
Eq.~\eqref{integral_I} can be written as a finite sum over one-dimensional integrals, 
\begin{multline}
\label{integral_I_kinds}
I_{\substack{n_{p_1},n_{p_2},\cdot\cdot\cdot,n_{p_N}\\m_{q_1},m_{q_2},\cdot\cdot\cdot,m_{q_N}}}^{j}=\frac{\Pi_{k=1}^{N}N_{p_k}N_{m_{q_k}}}{(j-1)!(N-j)!}
\sum_{\nu_1^{(1)}=0}^{\min(n_{p_1},m_{q_1})}\Bigg\{a_{\nu_1^{(1)}}\times\cdot\cdot\cdot\sum_{\nu_1^{(j-1)}=0}^{\min(n_{p_{j-1}},m_{q_{j-1}})}\Bigg\{a_{\nu_1^{(j-1)}}\times
\sum_{\nu_{1}^{(j+1)}=0}^{\min(n_{p_{j+1}},m_{q_{j+1}})}\Bigg\{a_{\nu_1^{(j+1)}}\times\\
\cdot\cdot\cdot\sum_{\nu_{1}^{(N)}=0}^{\min(n_{p_N},m_{q_N})}\Bigg\{a_{\nu_1^{(N)}}
 \mathcal{F}_{\substack{n_{p_1},n_{p_2},\cdot\cdot\cdot,n_{p_N}\\m_{q_1},m_{q_2},\cdot\cdot\cdot,m_{q_N}}}^{j}(\nu_1^{(1)},\cdot\cdot\cdot,\nu_1^{(j-1)},\nu_1^{(j+1)},\cdot\cdot\cdot,\nu_1^{(N)})\Bigg\}\cdot\cdot\cdot\Bigg\}\Bigg\}\cdot\cdot\cdot\Bigg\},
\end{multline}
where
\begin{multline}
\label{mathecal_f}
\mathcal{F}_{\substack{n_{p_1},n_{p_2},\cdot\cdot\cdot,n_{p_N}\\m_{q_1},m_{q_2},\cdot\cdot\cdot,m_{q_N}}}^{j}(\nu_1^{(1)},\cdot\cdot\cdot,\nu_1^{(j-1)},\nu_1^{(j+1)},\cdot\cdot\cdot,\nu_1^{(N)})=
\int_{-\infty}^{\infty}H_{n_{p_j}}(x_j)H_{m_{q_j}}(x_j)\exp(2ik_{so}x_j-x_j^2)\times\\
\Pi_{k<j}\left[
\left(\frac{\sqrt{\pi}}{2}-\frac{\sqrt{\pi}}{2}\text{erf}(x_j)\right)\delta_{n_{p_k}+m_{q_k}-2\nu_1^{(k)},0}
+\exp(-x_{j}^2)H_{n_{p_k}+m_{q_k}-2\nu_1^{(k)}-1}(x_{j})(1-\delta_{n_{p_k}+m_{q_k}-2\nu_1^{(k)},0})\right]\times\\
\Pi_{l>j}\left[\left(\frac{\sqrt{\pi}}{2}+\frac{\sqrt{\pi}}{2}\text{erf}(x_j)\right)\delta_{n_{p_l}+m_{q_l}-2\nu_1^{(l)},0}
-\exp(-x_{j}^2)H_{n_{p_l}+m_{q_l}-2\nu_1^{(l)}-1}(x_{j})(1-\delta_{n_{p_l}+m_{q_l}-2\nu_1^{(l)},0})\right]dx_j.
\end{multline}
To simplify the notation,
we define the index function $d(k)$,
\begin{equation}
\label{index_function}
d(k)=n_{p_k}+m_{q_k}-2\nu_1^{(k)}.
\end{equation}
For $k<j$ and $l>j$, we use $k_{1},k_{2},\cdot\cdot\cdot ,k_{\mathcal{K}}$ 
and $l_{1},l_{2},\cdot\cdot\cdot,l_{\mathcal{L}}$, respectively,
to indicate all the $k's$ and $l's$ that make $d(k)$ and $d(l)$ non-zero
($0\le \mathcal{K}\le j-1$ 
and 
$0\le \mathcal{L}\le N-j$).
Then Eq.~\eqref{mathecal_f} becomes
\begin{multline}
\label{simplify_mathecal_f}
\mathcal{F}_{\substack{n_{p_1},n_{p_2},\cdot\cdot\cdot,n_{p_N}\\m_{q_1},m_{q_2},\cdot\cdot\cdot,m_{q_N}}}^{j}(\nu_1^{(1)},\cdot\cdot\cdot,\nu_1^{(j-1)},\nu_1^{(j+1)},\cdot\cdot\cdot,\nu_1^{(N)})=\\
\int_{-\infty}^{\infty}H_{n_{p_j}}(x_j)H_{m_{q_j}}(x_j)
\exp\left[2ik_{so}x_j-(\mathcal{K}+\mathcal{L}+1)x_j^2\right]
(-1)^{\mathcal{L}}\left(\frac{\sqrt\pi}{2}-\frac{\sqrt\pi}{2}\text{erf}(x_j)\right)^{j-1-\mathcal{K}}\left(\frac{\sqrt\pi}{2}+\frac{\sqrt\pi}{2}\text{erf}(x_j)\right)^{N-j-\mathcal{L}}\times\\
H_{d(k_1)-1}(x_j)\cdot\cdot\cdot H_{d(k_{\mathcal{K}})-1}(x_j)H_{d(l_1)-1}(x_j)\cdot\cdot\cdot H_{d(l_{\mathcal{L}})-1}(x_j)dx_j.
\end{multline}
Expanding the $\left(\frac{\sqrt\pi}{2}-\frac{\sqrt\pi}{2}\text{erf}(x_j)\right)^{j-1-\mathcal{K}}$ and
$\left(\frac{\sqrt\pi}{2}+\frac{\sqrt\pi}{2}\text{erf}(x_j)\right)^{N-j-\mathcal{L}}$ terms,
Eq.~\eqref{simplify_mathecal_f} becomes
\begin{multline}
\label{simplify_mathecal_f_final}
\mathcal{F}_{\substack{n_{p_1},n_{p_2},\cdot\cdot\cdot,n_{p_N}\\m_{q_1},m_{q_2},\cdot\cdot\cdot,m_{q_N}}}^{j}(\nu_1^{(1)},\cdot\cdot\cdot\nu_1^{(j-1)},\nu_1^{(j+1)},\cdot\cdot\cdot\nu_1^{(N)})=\\
\left(\frac{\sqrt{\pi}}{2}\right)^{N-\mathcal{L}-\mathcal{K}-1}(-1)^{\mathcal{L}}\sum_{r=0}^{j-1-\mathcal{K}}\sum_{s=0}^{N-j-\mathcal{L}}\binom{j-1-\mathcal{K}}{r}
\binom{N-j-\mathcal{L}}{s}(-1)^{r}\times
\int_{-\infty}^{\infty}
\exp
\left[
2ik_{so}x_j-(\mathcal{K}+\mathcal{L}+1)x_j^2
\right]
\left[\text{erf}(x_j)\right]^{r+s}\times\\
H_{n_{p_j}}(x_j)H_{m_{q_j}}(x_j)H_{d(k_1)-1}(x_j)
\cdot\cdot\cdot H_{d(k_{\mathcal{K}})-1}(x_j)H_{d(l_1)-1}(x_j)\cdot\cdot\cdot H_{d(l_{\mathcal{L}})-1}(x_j)dx_j.
\end{multline}
Equation~\eqref{simplify_mathecal_f_final} 
contains a product of Hermite polynomials,
which can be converted into a finite sum of Hermite polynomials according to the 
Feldheim identity,
\begin{equation}
\label{expand_Hermite}
H_{n_{p_j}}(x_j)H_{m_{q_j}}(x_j)H_{d(k_1)-1}(x_j)
\cdot\cdot\cdot H_{d(k_{\mathcal{K}})-1}(x_j)H_{d(l_1)-1}(x_j)\cdot\cdot\cdot H_{d(l_{\mathcal{L}})-1}(x_j)=
\sum_{\nu_1\cdot\cdot\cdot\nu_{\mathcal{L}+\mathcal{K}+1}}a_{\nu_1\cdot\cdot\cdot\nu_{\mathcal{L}+\mathcal{K}+1}}H_{M}(x_j),
\end{equation}
where the coefficients $a_{\nu_1\cdot\cdot\cdot\nu_{\mathcal{L}+\mathcal{K}+1}}$ and the relationship 
between $M$ and 
the indices $n_{p_j}, m_{q_j},d(k_1),\cdot\cdot\cdot, d(l_{\mathcal{L}})$ are 
defined in Eqs.~\eqref{M}-\eqref{limit}.

Plugging Eq.~\eqref{expand_Hermite} into 
Eq.~\eqref{simplify_mathecal_f_final}, 
we find
\begin{multline}
\label{simplify_mathecal_f_final_final}
\mathcal{F}_{\substack{n_{p_1},n_{p_2},\cdot\cdot\cdot,n_{p_N}\\m_{q_1},m_{q_2},\cdot\cdot\cdot,m_{q_N}}}^{j}(\nu_1^{(1)},\cdot\cdot\cdot,\nu_1^{(j-1)},\nu_1^{(j+1)},\cdot\cdot\cdot,\nu_1^{(N)})=\\
\left(\frac{\sqrt{\pi}}{2}\right)^{N-\mathcal{L}-\mathcal{K}-1}(-1)^{\mathcal{L}}\sum_{r=0}^{j-1-\mathcal{K}}\sum_{s=0}^{N-j-\mathcal{L}}\binom{j-1-\mathcal{K}}{r}
\binom{N-j-\mathcal{L}}{s}(-1)^{r}
\sum_{\nu_1\cdot\cdot\cdot\nu_{\mathcal{L}+\mathcal{K}+1}}a_{\nu_1\cdot\cdot\cdot\nu_{\mathcal{L}+\mathcal{K}+1}}\mathcal{G}^{M}_{(r+s;\mathcal{K}+\mathcal{L})},
\end{multline}
where
\begin{equation}
\label{mathecal_g}
\mathcal{G}_{(r+s;\mathcal{K}+\mathcal{L})}^{M}=\int_{-\infty}^{\infty}\exp\left[2ik_{so}x_j-(\mathcal{K}+\mathcal{L}+1)x_j^2\right]\left[\text{erf}(x_j)\right]^{r+s}H_M(x_j)dx_j.
\end{equation}
We call $\mathcal{G}_{(r+s;\mathcal{K}+\mathcal{L})}^{M}$ the $\mathcal{G}$ integral 
of the order $r+s$.
The $\mathcal{G}$ integral of order $0$ can be evaluated analytically,
\begin{equation}
\label{mathecal_g_zero_order}
\mathcal{G}_{(0,\mathcal{K}+\mathcal{L})}^{M}
=\frac{\sqrt{\pi}}{\sqrt{\mathcal{K}+\mathcal{L}+1}}\exp\left(-\frac{k_{so}^2}{\mathcal{K}+\mathcal{L}+1}\right)
H_{M}\left(\frac{2ik_{so}}{\sqrt{(\mathcal{K}+\mathcal{L}+1)(\mathcal{K}+\mathcal{L})}}\right)\left(\frac{\mathcal{K}+\mathcal{L}}{\mathcal{K}+\mathcal{L}+1}\right)^{M/2}.
\end{equation}
To evaluate $\mathcal{G}^{M}_{(r+s;\mathcal{K}+\mathcal{L})}$ 
of higher order, 
we develop an iterative procedure, in which the $\mathcal{G}$ integral of order
$r+s$ is written in terms of $\mathcal{G}$ integrals of order $r+s-1$. Since
the $r+s=0$ integral 
is known, this allows for the evaluation of the $\mathcal{G}$ integral
of arbitrary order.
Using the generating function of the Hermite polynomials,
we have
\begin{equation}
\label{generating_mathecal_g}
\mathcal{G}_{(r+s;\mathcal{K}+\mathcal{L})}^{M}=\left\{\frac{\partial^M}{\partial t^M}\int_{-\infty}^{\infty}\exp\left[2(ik_{so}+t)x_j-(\mathcal{K}+\mathcal{L}+1)x_j^2-t^2\right]
\left[\text{erf}(x_j)\right]^{r+s}dx_j\right\}
\Bigg|_{t=0}.
\end{equation}
Using integration by parts, Eq.~\eqref{generating_mathecal_g} becomes
\begin{equation}
\label{g_integral_by_parts}
\mathcal{G}_{(r+s;\mathcal{K}+\mathcal{L})}^{M}=
\Bigg\{\frac{\partial^M}{\partial t^M}
\Bigg[
g_{(r+s,\mathcal{K}+\mathcal{L})}^{(1)}-
g_{(r+s,\mathcal{K}+\mathcal{L})}^{(2)}
\Bigg] \Bigg\}
\Bigg|_{t=0},
\end{equation}
where
\begin{equation}
\label{g1}
g_{(r+s,\mathcal{K}+\mathcal{L})}^{(1)}
=
\mathcal{H}(k_{so},\mathcal{K},\mathcal{L};t,x_j)\left[\text{erf}(x_j)\right]^{r+s}|_{x_j=-\infty}^{x_j=\infty}
\end{equation}
and
\begin{equation}
\label{g2}
g_{(r+s,\mathcal{K}+\mathcal{L})}^{(2)}
=
\left(r+s\right)\frac{2}{\sqrt{\pi}}\int_{-\infty}^{\infty}\mathcal{H}(k_{so},\mathcal{K},\mathcal{L};t,x_j)\left[\text{erf}(x_j)\right]^{r+s-1}
\exp(-x_j^2)
dx_j
\end{equation}
with
\begin{equation}
\label{mathecal_H}
\mathcal{H}(k_{so},\mathcal{K},\mathcal{L};t,x_j)=\int_{0}^{x_j} \exp\left[2(ik_{so}+t)x-(\mathcal{K}+\mathcal{L}+1)x^2-t^2\right]dx.
\end{equation}
The Gaussian
integral in Eq.~\eqref{mathecal_H} 
can be calculated analytically,
\begin{equation}
\label{mathecal_H_final}
\mathcal{H}(k_{so},\mathcal{K},\mathcal{L};t,x_j)
=
\frac{\sqrt{\pi}
}{2\sqrt{\mathcal{K}+\mathcal{L}+1}}
\exp\left[-\frac{k_{so}^2-2ik_{so}t+(\mathcal{K}+\mathcal{L})t^2}{\mathcal{K}+\mathcal{L}+1}\right]
\text{erf}\left[\frac{-ik_{so}-t+(\mathcal{K}+\mathcal{L}+1)x_j}{\sqrt{\mathcal{K}+\mathcal{L}+1}}\right].
\end{equation}
Using 
that the value of the error function $\text{erf}(x)$ at infinity is known, $\text{erf}(\pm\infty)=\pm 1$,
$g^{(1)}_{(r+s,\mathcal{K}+\mathcal{L})}$ evaluates to
\begin{equation}
\label{g_first_term}
g_{(r+s,\mathcal{K}+\mathcal{L})}^{(1)}
=
\frac{\sqrt{\pi}}{2\sqrt{\mathcal{K}+\mathcal{L}+1}}
\exp\left[-\frac{k_{so}^2-2ik_{so}t+(\mathcal{K}+\mathcal{L})t^2}{\mathcal{K}+\mathcal{L}+1}\right]
\left[1-(-1)^{r+s+1}\right]
.
\end{equation}
Inserting Eq.~(\ref{mathecal_H_final}) into Eq.~(\ref{g2}), we have
\begin{equation}
\label{g2_final}
g_{(r+s,\mathcal{K}+\mathcal{L})}^{(2)}=(r+s)\frac{\exp\left[-\frac{k_{so}^2-2ik_{so}t+(\mathcal{K}+\mathcal{L})t^2}{\mathcal{K}+\mathcal{L}+1}\right]}{\sqrt{\mathcal{K}+\mathcal{L}+1}}
\left\{\int_{-\infty}^{\infty}\text{erf}\left[\frac{-ik_{so}-t+(\mathcal{K}+\mathcal{L}+1)x_j}{\sqrt{\mathcal{K}+\mathcal{L}+1}}\right]\left[\text{erf}(x_j)\right]^{r+s-1}\exp(-x_j^2)dx_j\right\}.
\end{equation}

Next, we expand
$g_{(r+s,\mathcal{K}+\mathcal{L})}^{(1)}$ and 
$g_{(r+s,\mathcal{K}+\mathcal{L})}^{(2)}$ in terms of $t$
up to power $M$ and evaluate the $M$th derivative with respect to $t$ at $t=0$.
For $\mathcal{K}+\mathcal{L}\neq 0$, 
the exponential in Eq.~(\ref{g_first_term}) can be interpreted as 
a generating function of the 
Hermite polynomials.
Expanding this term into a sum of 
Hermite polynomials, we find
\begin{multline}
\label{mathecal_g_expand_t_first_term_nonzero}
g_{(r+s,\mathcal{K}+\mathcal{L})}^{(1)}=\\
\frac{\sqrt{\pi}}{2\sqrt{\mathcal{K}+\mathcal{L}+1}}\left[1-(-1)^{r+s+1}\right]
\exp\left[-\frac{k_{so}^2}{\mathcal{K}+\mathcal{L}+1}+\frac{2ik_{so}}{\sqrt{(\mathcal{K}+\mathcal{L}+1)(\mathcal{K}+\mathcal{L})}}
\left(\sqrt{\frac{\mathcal{K}+\mathcal{L}}{\mathcal{K}+\mathcal{L}+1}}t\right)-\left(\sqrt{\frac{\mathcal{K}+\mathcal{L}}{\mathcal{K}+\mathcal{L}+1}}t\right)^2\right]\\
=\frac{\sqrt{\pi}}{2\sqrt{\mathcal{K}+\mathcal{L}+1}}\left[1-(-1)^{r+s+1}\right]
\exp\left(-\frac{k_{so}^2}{\mathcal{K}+\mathcal{L}+1}\right)
\left[
\sum_{n=0}^{\infty}
H_{n}\left(\frac{2ik_{so}}{\sqrt{(\mathcal{K}+\mathcal{L}+1)(\mathcal{K}+\mathcal{L})}}\right)\frac{\left(\sqrt{\frac{\mathcal{K}+\mathcal{L}}{\mathcal{K}+\mathcal{L}+1}}t\right)^n}{n!}
\right].
\end{multline} 
For $\mathcal{K}+\mathcal{L}=0$, the power 
series
of $g_{(r+s,\mathcal{K}+\mathcal{L})}^{(1)}$ in terms of $t$ is
\begin{equation}
\label{mathecal_g_expand_t_first_term_zero}
g_{(r+s,\mathcal{K}+\mathcal{L})}^{(1)}=\frac{\sqrt{\pi}}{2\sqrt{\mathcal{K}+\mathcal{L}+1}}\left[1-(-1)^{r+s+1}\right]\exp\left(-k_{so}^2\right)\left[\sum_{n=0}^{\infty}\frac{(2ik_{so}t)^n}{n!}\right].
\end{equation} 
This can
be understood as the limiting result of Eq.~\eqref{mathecal_g_expand_t_first_term_nonzero},
\begin{equation}
\label{limit_hermite}
\lim_{\mathcal{K}+\mathcal{L}\to 0}H_{n}\left(\frac{2ik_{so}}{\sqrt{(\mathcal{K}+\mathcal{L}+1)(\mathcal{K}+\mathcal{L})}}\right)
\left(\frac{\mathcal{K}+\mathcal{L}}{\mathcal{K}+\mathcal{L}+1}\right)^{n/2}=(2ik_{so})^n.
\end{equation}
In what follows, we take this limit 
when
$\mathcal{K}+\mathcal{L}=0$.
The derivative of $g^{(1)}_{(r+s;\mathcal{K}+\mathcal{L})}$ with respect to $t$ 
at $t=0$ is then
\begin{multline}
\label{g1_derivative}
\left(
\frac{\partial^M}{\partial t^M}g_{(r+s;\mathcal{K}+\mathcal{L})}^{(1)}
\right)
\Bigg|_{t=0}=\\
\frac{\sqrt{\pi}}{2\sqrt{\mathcal{K}+\mathcal{L}+1}}
\exp\left(-\frac{k_{so}^2}{\mathcal{K}+\mathcal{L}+1}\right)
H_{M}\left(\frac{2ik_{so}}{\sqrt{(\mathcal{K}+\mathcal{L}+1)(\mathcal{K}+\mathcal{L})}}\right)\left(\frac{\mathcal{K}+\mathcal{L}}{\mathcal{K}+\mathcal{L}+1}\right)^{M/2}
\left[1-(-1)^{r+s+1}\right].
\end{multline}

To evaluate $g_{(r+s,\mathcal{K}+\mathcal{L})}^{(2)}$, we notice that it 
can be written in the form 
$g_{(r+s,\mathcal{K}+\mathcal{L})}^{(1)} \int\text{erf}(\cdots) dx_j$.
The expansion of $g_{(r+s,\mathcal{K}+\mathcal{L})}^{(1)}$ is given in
Eq.~\eqref{mathecal_g_expand_t_first_term_nonzero}.
An additional $t$-dependence enters through the error function in the
integrand in Eq.~\eqref{g2_final}. 
Rewriting the error function
in a series in $t$, we have
\begin{multline}
\label{mathecal_g_expand_t_second_term}
\text{erf}\Bigg(\frac{-ik_{so}-t+(\mathcal{K}+\mathcal{L}+1)x_j}{\sqrt{\mathcal{K}+\mathcal{L}+1}}\Bigg)=
\text{erf}\Bigg(\sqrt{\mathcal{K}+\mathcal{L}+1}x_j-\frac{ik_{so}}{\sqrt{\mathcal{K}+\mathcal{L}+1}}\Bigg)-
\frac{2}{\sqrt{\pi}}\exp\left(\frac{k_{so}^2}{\mathcal{K}+\mathcal{L}+1}\right)\times\\
\sum_{n=1}^{\infty}\exp\left[
2ik_{so}x_j-(\mathcal{K}+\mathcal{L}+1)x_j^2
\right]H_n\left(\sqrt{\mathcal{K}+\mathcal{L}+1}x_j-\frac{ik_{so}}{\sqrt{\mathcal{K}+\mathcal{L}+1}}\right)
\Bigg(\frac{t}{\sqrt{\mathcal{K}+\mathcal{L}+1}}\Bigg)^n\frac{1}{n!}.
\end{multline}
Using the multiplication theorem~\cite{multiplication_theorem}
\begin{equation}
\label{formula1}
H_{n}(\alpha x)=\sum_{v=0}^{[\frac{n}{2}]}\alpha^{n-2v}(\alpha^2-1)^v\binom{n}{2v}\frac{(2v)!}{v!}H_{n-2v}(x)
\end{equation}
and the addition theorem~\cite{addition_theorem}
\begin{equation}
\label{formula2}
H_n(x+y)=\sum_{w=0}^{n}\binom{n}{w}H_{w}(x)(2y)^{n-w},
\end{equation}
where $[\frac{n}{2}]$ denotes the integer part of $n/2$,
the Hermite polynomial on the right hand side of Eq.~\eqref{mathecal_g_expand_t_second_term} 
can be
expanded into a finite sum 
over products
of Hermite polynomials in which the dependence on $x_j$ has been 
``isolated'',
\begin{multline}
\label{expansion_hermite_mathecal_g_expand_t_second_term}
H_n\left(\sqrt{\mathcal{K}+\mathcal{L}+1}x_j-\frac{ik_{so}}{\sqrt{\mathcal{K}+\mathcal{L}+1}}\right)=\\
\sum_{v=0}^{[\frac{n}{2}]}\sum_{w=0}^{n-2v}\left(\mathcal{K}+\mathcal{L}+1\right)^{\frac{n-2v}{2}}\left(\mathcal{K}+\mathcal{L}\right)^v
\binom{n}{2v}\frac{(2v)!}{v!}\binom{n-2v}{w}H_{w}
\left( 
\frac{2ik_{so}}{\mathcal{K}+\mathcal{L}+1}
\right)
H_{n-2v-w}(x_j).
\end{multline}
For $\mathcal{K}+\mathcal{L}=0$,
Eq.~\eqref{expansion_hermite_mathecal_g_expand_t_second_term} 
contains an indeterminate term $0^0$ which is understood to be $1$.
In this case, Eq.~\eqref{expansion_hermite_mathecal_g_expand_t_second_term} 
becomes
\begin{equation}
\label{expansion_hermite_mathecal_g_expand_t_second_term_zero}
H_n\left(x_j-ik_{so}\right)=
\sum_{w=0}^{n}
\binom{n}{w}H_{w}
\left(2ik_{so}
\right)
H_{n-w}(x_j).
\end{equation}

Using 
Eqs.~\eqref{mathecal_g_expand_t_first_term_nonzero}, 
~\eqref{mathecal_g_expand_t_second_term}
and 
~\eqref{expansion_hermite_mathecal_g_expand_t_second_term},
Eq.~\eqref{g2_final}
becomes
\begin{multline}
\label{expansion_g2}
g_{(r+s;\mathcal{K}+\mathcal{L})}^{(2)}=\\
\sum_{n=0}^{\infty}\Bigg\{\frac{r+s}{\sqrt{\mathcal{K}+\mathcal{L}+1}}\exp\left(-\frac{k_{so}^2}{\mathcal{K}+\mathcal{L}+1}\right)\Bigg[
H_{n}\left(\frac{2ik_{so}}{\sqrt{(\mathcal{K}+\mathcal{L}+1)(\mathcal{K}+\mathcal{L})}}\right)\left(\frac{\mathcal{K}+\mathcal{L}}{\mathcal{K}+\mathcal{L}+1}\right)^{n/2}
\mathcal{I}_{(r+s-1,\mathcal{K}+\mathcal{L})}\Bigg]+\\
\frac{2(r+s)}{\sqrt{\pi}(\mathcal{K}+\mathcal{L}+1)^{\frac{n+1}{2}}}\sum_{u=0}^{n}\sum_{v=0}^{[\frac{u}{2}]}\sum_{w=0}^{u-2v}
H_{n-u}\left(\frac{2ik_{so}}{\sqrt{(\mathcal{K}+\mathcal{L}+1)(\mathcal{K}+\mathcal{L})}}\right)
\left(\mathcal{K}+\mathcal{L}\right)^{\frac{n-u}{2}+v}
\left(\mathcal{K}+\mathcal{L}+1\right)^{\frac{u-2v}{2}}\times\\
\binom{n}{u}\binom{u}{2v}\binom{u-2v}{w}\frac{(2v)!}{v!}H_{w}\left(\frac{2ik_{so}}{\mathcal{K}+\mathcal{L}+1}\right)
\mathcal{G}_{(r+s-1,\mathcal{K}+\mathcal{L}+1)}^{(u-2v-w)}\Bigg\}\frac{t^n}{n!},
\end{multline}
where
\begin{equation}
\label{mathecal_I}
\mathcal{I}_{(r+s-1,\mathcal{K}+\mathcal{L})}=\int_{-\infty}^{\infty}\text{erf}\Bigg(\sqrt{\mathcal{K}+\mathcal{L}+1}x_j-\frac{ik_{so}}{\sqrt{\mathcal{K}+\mathcal{L}+1}}\Bigg)
\left[\text{erf}(x_j)\right]^{r+s-1}\exp(-x_j^2)dx_j.
\end{equation}
The $M$th
derivative of $g_{(r+s;\mathcal{K}+\mathcal{L})}^{(2)}$ with respect to $t$ 
at $t=0$ is then
\begin{equation}
\label{g2_derivative}
\left(
\frac{\partial^M}{\partial t^M}g_{(r+s;\mathcal{K}+\mathcal{L})}^{(2)}
\right) 
\Bigg|_{t=0}=
g_{(r+s,\mathcal{K}+\mathcal{L})}^{(2,1)}
+
g_{(r+s,\mathcal{K}+\mathcal{L})}^{(2,2)},
\end{equation}
where
\begin{multline}
\label{g21}
g_{(r+s,\mathcal{K}+\mathcal{L})}^{(2,1)}
= \\
\frac{r+s}{\sqrt{\mathcal{K}+\mathcal{L}+1}}\exp\left(-\frac{k_{so}^2}{\mathcal{K}+\mathcal{L}+1}\right)
H_{M}\left(\frac{2ik_{so}}{\sqrt{(\mathcal{K}+\mathcal{L}+1)(\mathcal{K}+\mathcal{L})}}\right)\left(\frac{\mathcal{K}+\mathcal{L}}{\mathcal{K}+\mathcal{L}+1}\right)^{M/2}
\mathcal{I}_{(r+s-1,\mathcal{K}+\mathcal{L})}
\end{multline}
and
\begin{multline}
\label{g22}
g_{(r+s,\mathcal{K}+\mathcal{L})}^{(2,2)}
=
\frac{2(r+s)}{\sqrt{\pi}(\mathcal{K}+\mathcal{L}+1)^{\frac{M+1}{2}}}\sum_{u=0}^{M}\sum_{v=0}^{[\frac{u}{2}]}\sum_{w=0}^{u-2v}
H_{M-u}\left(\frac{2ik_{so}}{\sqrt{(\mathcal{K}+\mathcal{L}+1)(\mathcal{K}+\mathcal{L})}}\right)
\left(\mathcal{K}+\mathcal{L}\right)^{\frac{M-u}{2}+v}
\left(\mathcal{K}+\mathcal{L}+1\right)^{\frac{u-2v}{2}}\times\\
\binom{M}{u}\binom{u}{2v}\binom{u-2v}{w}\frac{(2v)!}{v!}H_{w}\left(\frac{2ik_{so}}{\mathcal{K}+\mathcal{L}+1}\right)
\mathcal{G}_{(r+s-1,\mathcal{K}+\mathcal{L}+1)}^{(u-2v-w)}.
\end{multline}
To
evaluate the integral $\mathcal{I}_{(r+s-1,\mathcal{K}+\mathcal{L})}$,
we integrate by parts.
Using 
\begin{equation}
\label{err_integral_by_part}
\int_{0}^{x_j}\left[\text{erf}(x)\right]^{r+s-1}\exp(-x^2)dx=\frac{\sqrt{\pi}\left[\text{erf}(x_j)\right]^{r+s}}{2(r+s)}
\end{equation}
and
\begin{equation}
\label{mathecal_I_first_part}
\left[
\text{erf}\Bigg(\sqrt{\mathcal{K}+\mathcal{L}+1}x_j-\frac{ik_{so}}{\sqrt{\mathcal{K}+\mathcal{L}+1}}\Bigg)
\frac{\sqrt{\pi}\left[\text{erf}(x_j)\right]^{r+s}}{2(r+s)}
\right]
\bigg|_{x_j=-\infty}^{x_j=\infty}=
\frac{\sqrt{\pi}}{2(r+s)}
\left[
1-(-1)^{r+s+1}
\right],
\end{equation} 
we find
\begin{equation}
\label{mathecal_I_by_part}
\mathcal{I}_{(r+s-1,\mathcal{K}+\mathcal{L})}=
\frac{\sqrt{\pi}}{2(r+s)}
\left[
1-(-1)^{r+s+1}
\right]
-\frac{\sqrt{\mathcal{K}+\mathcal{L}+1}\exp\left(\frac{k_{so}^2}{\mathcal{K}+\mathcal{L}+1}\right)}{r+s}\mathcal{J}_{(r+s,\mathcal{K}+\mathcal{L})},
\end{equation}
where
\begin{equation}
\label{mathecal_J}
\mathcal{J}_{(r+s,\mathcal{K}+\mathcal{L})}=\int_{-\infty}^{\infty}\left[\text{erf}(x_j)\right]^{r+s}\exp[-(\mathcal{K}+\mathcal{L}+1)x_j^2+2ik_{so}x_j]dx_j.
\end{equation}
Using Eq.~\eqref{mathecal_I_by_part} in Eq.~\eqref{g21} and using Eq.~\eqref{g1_derivative},
we find
\begin{equation}
\label{mathecal_g_final_final}
\mathcal{G}_{(r+s;\mathcal{K}+\mathcal{L})}^{M}=
\mathcal{J}_{(r+s,\mathcal{K}+\mathcal{L})}-
g_{(r+s,\mathcal{K}+\mathcal{L})}^{(2,2)}.
\end{equation}
Our manipulations have reduced 
$\mathcal{G}_{(r+s;\mathcal{K}+\mathcal{L})}^{M}$ to an expression that
contains two types of one-dimensional integrals.
The first one, $\mathcal{J}_{(r+s,\mathcal{K}+\mathcal{L})}$,
only depends on $k_{so}$,
$\mathcal{K}$ and $\mathcal{L}$,
implying that only a few of these integrals need to be calculated for 
a given $k_{so}$. 
For $r+s=1$, $\mathcal{J}_{(1,\mathcal{K}+\mathcal{L})}$ can be calculated analytically,
\begin{equation}
\label{mathecal_J_analytical}
\mathcal{J}_{(1,\mathcal{K}+\mathcal{L})}=\sqrt{\frac{\pi}{\mathcal{K}+\mathcal{L}+1}}\exp\left(-\frac{k_{so}^2}{\mathcal{K}+\mathcal{L}+1}\right)
\text{erf}\left[\frac{ik_{so}}{\sqrt{(\mathcal{K}+\mathcal{L}+1)(\mathcal{K}+\mathcal{L}+2)}}\right].
\end{equation}
For $r+s>1$, $\mathcal{J}_{(1,\mathcal{K}+\mathcal{L})}$ can be calculated numerically
with essentially arbitrary accuracy.
The second 
integral, $\mathcal{G}^{u-2v-w}_{(r+s-1;\mathcal{K}+\mathcal{L}+1)}$
(the $\mathcal{G}$ integral of order $r+s-1$),  is
contained in 
$g_{(r+s,\mathcal{K}+\mathcal{L})}^{(2,2)}$. 
To see the iterative structure of our result more clearly, we rewrite
Eq.~\eqref{mathecal_g_final_final} as
\begin{equation}
\mathcal{G}_{(r+s;\mathcal{K}+\mathcal{L})}^{M}=
\mathcal{J}_{(r+s,\mathcal{K}+\mathcal{L})}-
\sum_j c_j
\mathcal{G}^{u-2v-w}_{(r+s-1;\mathcal{K}+\mathcal{L}+1)},
\end{equation}
where $j$ runs over all combinations of allowed indices and $c_j$ contains
all the prefactors [the $c_j$ can be read off Eq.~\eqref{g22}].
Thus,
to determine the $\mathcal{G}$ integral of order $r+s$,
a finite number of  
$\mathcal{G}$ integrals of order $r+s-1$ is needed.
To evaluate the $\mathcal{G}$ integral of order $r+s-1$, 
a finite number of 
$\mathcal{G}$ integrals of order $r+s-2$
is needed, and so on.
Since the $\mathcal{G}$ integral
of order $0$ is known analytically [see Eq.~\eqref{mathecal_g_zero_order}],
the $\mathcal{G}$ integral of arbitrary order can be obtained iteratively.
It should be noted that the $\mathcal{G}$ integral of order $1$
has an analytical result since $\mathcal{G}_{(0,\mathcal{K}+\mathcal{L})}$
and $\mathcal{J}_{(1,\mathcal{K}+\mathcal{L})}$ are known analytically.

In summary, first we calculate,
using Eq.~\eqref{mathecal_J},
all the $\mathcal{J}$ integrals 
needed for the evaluation of the $\mathcal{G}$ integrals. 
Secondly, using the values of the $\mathcal{J}$ integrals, 
the required
$\mathcal{G}$ integrals 
are calculated iteratively based on Eq.~\eqref{mathecal_g_final_final}.
Thirdly, plugging the values of the $\mathcal{G}$ integrals into Eq.~\eqref{simplify_mathecal_f_final_final}, 
the $\mathcal{F}$ integrals with arguments $\nu_1^{(1)},\cdot\cdot\cdot,\nu_1^{(j-1)},\nu_1^{(j+1)},\cdot\cdot\cdot,\nu_1^{(N)}$ are obtained.
Forthly, plugging the values of the $\mathcal{F}$ integrals into Eq.~\eqref{integral_I_kinds}, the $I$ integrals can be calculated.
Finally, plugging the values of the $I$ integrals into Eq.~\eqref{double_sum}, we 
obtain
the values of the $\mathcal{D}$ integrals.
The sums in the 
above expressions are all finite.
As a result, 
the only error in evaluating the $\mathcal{D}$ integrals
comes from the numerical evaluation of the 
one-dimensional $\mathcal{J}$ integrals.
Everything is analytical for $N=2$
while
one and three $\mathcal{J}$ integrals have to 
be evaluated numerically
for $N=3$ and $4$, respectively.
The reader may wonder why we choose the outlined iterative 
procedure for evaluating the one-dimensional integral
given in Eq.~\eqref{mathecal_g}
over an evaluation of
Eq.~\eqref{mathecal_g} by direct numerical integration.
The answer is two-fold. 
First, Eq.~\eqref{mathecal_g} has to be evaluated for Hermite
polynomials of different orders; the numerical
integrals required in our iterative
procedure, in contrast, are independent of the
$M$ quantum number.
Second,
for large $M$, the integrand in Eq.~\eqref{mathecal_g} is
highly oscillatory, making the one-dimensional
integration somewhat non-trivial.

To be concrete, we 
discuss
how to apply the 
formalism
to  
the 
three particle system.
In this case,
the $\mathcal{D}$ integral depends 
on $j$ ($j=1,2,3$) and the quantum numbers 
$n_1$, $n_2$, $n_3$, $m_1$, $m_2,$ and $m_3$.
For a 
fixed $j$, there are $3!\times3!=36$ terms
($I$ integrals) in the sum on the right hand side of Eq.~\eqref{double_sum}. 
For each $I$ integral, we have a double sum on the right hand side of Eq.~\eqref{integral_I_kinds}.
For $j=2$, e.g., the summation dummies are
$\nu_1^{(1)}$ and $\nu_1^{(3)}$.
Correspondingly, the $\mathcal{F}$ integrals have two arguments.
The indices $\mathcal{K}$ and $\mathcal{L}$ in Eq.~\eqref{simplify_mathecal_f} satisfy 
$0\le \mathcal{K}\le 1$
and
$0\le \mathcal{L}\le 1$.
As a result, the indices $r$ and $s$ defined in Eq.~\eqref{simplify_mathecal_f_final} satisfy 
$0\le r\le 1$
and
$0\le s\le 1$ with the additional constraint 
$r+s\le N-1-\mathcal{K}-\mathcal{L},$
which is due to the limits 
$r\le j-\mathcal{K}-1$ and 
$s\le N-\mathcal{L}-j$ 
in the sums on the right hand side of Eq.~\eqref{simplify_mathecal_f_final}.
Thus, the allowed $(r+s,\mathcal{K}+\mathcal{L})$
combinations are $(2,0),(1,0),(0,0),(1,1),(0,1),(0,2)$.
Since the number of Hermite polynomials in the product in Eq.~\eqref{simplify_mathecal_f_final} is 
$\mathcal{K}+\mathcal{L}+2,$ 
which is less or equal to $4$
(since $0\le\mathcal{K}\le 1$ and $0\le\mathcal{L}\le 1$), 
the coefficients on the right hand side of Eq.~\eqref{expand_Hermite} have at most three indices.

\section{Expressions for $\langle S_x(x)\rangle$ and $\langle S_z(x)\rangle$ for $N=3-4$}
\label{appendix_B}
For $N=3$,
the expressions for $C_{1x}, C_{2x},$ and $C_z$ in Eqs.~\eqref{Sx_3particle}-\eqref{Sz_3particle}
in terms of the coefficients $C_1-C_4$ in Eq.~\eqref{eigenstate_3particle} are
\begin{equation}
C_{1x}=(C_{1}-C_{3})^{*}(C_{2}+C_{4})+(C_{1}-C_{3})(C_{2}+C_{4})^{*},
\end{equation}
\begin{equation}
C_{2x}=C_{1}^{*}C_{3}+C_{1}C_{3}^{*}-|C_{2}|^{2}-|C_{4}|^{2},
\end{equation}
and
\begin{equation}
C_{z}=i[(C_{1}-C_{3})^{*}(C_{2}-C_{4})-(C_{1}-C_{3})(C_{2}-C_{4})^{*}].
\end{equation}
The expressions for $n_{1x}(x), n_{2x}(x),$ and $n_z(x)$ in Eqs.~\eqref{Sx_3particle}-\eqref{Sz_3particle} are
\begin{equation}
n_{1x}(x)=\int_{-\infty}^{\infty}|D(0,1,2)|^{2}\Theta_{x_1<x_2<x_3}[\delta(x-x_{1})+\delta(x-x_{3})]d\vec{x},
\end{equation}
\begin{equation}
n_{2x}(x)=\int_{-\infty}^{\infty}|D(0,1,2)|^{2}\Theta_{x_1<x_2<x_3}\delta(x-x_{2})d\vec{x} ,
\end{equation}
and
\begin{equation}
n_{z}(x)=\int_{-\infty}^{\infty}|D(0,1,2)|^{2}\Theta_{x_1<x_2<x_3}[\delta(x-x_{1})-\delta(x-x_{3})]d\vec{x}.
\end{equation}

For $N=4$, 
the expressions for $C_{1x}, C_{2x}, C_{1z},$ and $C_{2z}$ in Eqs.~\eqref{Sx_4particle}-\eqref{Sz_4particle}
in terms of the coefficients $C_1-C_{10}$ in Eq.~\eqref{eigenstate_4particle} are
\begin{multline}
C_{1x}=\frac{1}{2}\left[\left(C_2+C_5\right)\left(C_1+C_8\right)^{*}+\left(C_2+C_5\right)^{*}\left(C_1+C_8\right)\right]\\
+\frac{1}{\sqrt{2}}\left[C_3\left(C_{10}+C_{6}\right)^{*}+C_3^{*}\left(C_{10}+C_{6}\right)
+C_4\left(C_7+C_9\right)^{*}+C_4^{*}\left(C_7+C_9\right)\right],
\end{multline}
\begin{multline}
C_{2x}=\frac{1}{2}\left[\left(C_3+C_4\right)\left(C_1+C_8\right)^{*}+\left(C_3+C_4\right)^{*}\left(C_1+C_8\right)\right]\\
+\frac{1}{\sqrt{2}}\left[C_5\left(C_{9}+C_{10}\right)^{*}+C_5^{*}\left(C_{9}+C_{10}\right)
+C_2\left(C_6+C_7\right)^{*}+C_2^{*}\left(C_6+C_7\right)\right],
\end{multline}
\begin{multline}
C_{1z}=\frac{i}{2}\left[\left(C_2-C_5\right)\left(C_1+C_8\right)^{*}-\left(C_2-C_5\right)^{*}\left(C_1+C_8\right)\right]\\
+\frac{i}{\sqrt{2}}\left[C_3\left(C_{10}-C_{6}\right)^{*}-C_3^{*}\left(C_{10}-C_{6}\right)
+C_4\left(C_9-C_7\right)^{*}-C_4^{*}\left(C_9-C_7\right)\right],
\end{multline}
and 
\begin{multline}
C_{2z}=\frac{i}{2}\left[\left(C_3-C_4\right)\left(C_1+C_8\right)^{*}-\left(C_3-C_4\right)^{*}\left(C_1+C_8\right)\right]\\
+\frac{i}{\sqrt{2}}\left[C_5\left(C_{9}-C_{10}\right)^{*}-C_5^{*}\left(C_{9}-C_{10}\right)
+C_2\left(C_7-C_6\right)^{*}-C_2^{*}\left(C_7-C_6\right)\right].
\end{multline}
The expressions for $n_{1x}(x), n_{2x}(x), n_{1z}(x),$ and $n_{2z}(x)$ in Eqs.~\eqref{Sx_4particle}-\eqref{Sz_4particle} are
\begin{equation}
n_{1x}(x)=\int_{-\infty}^{\infty}|D(0,1,2,4)|^{2}\Theta_{x_1<x_2<x_3<x_4}\left[\delta(x-x_{1})+\delta(x-x_4)\right]d\vec{x},
\end{equation}
\begin{equation}
n_{2x}(x)=\int_{-\infty}^{\infty}|D(0,1,2,4)|^{2}\Theta_{x_1<x_2<x_3<x_4}[\delta(x-x_{2})+\delta(x-x_{3})]d\vec{x},
\end{equation}
\begin{equation}
n_{1z}(x)=\int_{-\infty}^{\infty}|D(0,1,2,4)|^{2}\Theta_{x_1<x_2<x_3<x_4}[\delta(x-x_{1})-\delta(x-x_{4})]d\vec{x},
\end{equation}
and
\begin{equation}
n_{2z}(x)=\int_{-\infty}^{\infty}|D(0,1,2,4)|^{2}\Theta_{x_1<x_2<x_3<x_4}[\delta(x-x_{2})-\delta(x-x_{3})]d\vec{x}.
\end{equation}

\end{widetext}

\end{document}